\documentclass[twocolumn]{bmcart}% uncomment this for twocolumn layout and comment line below
\usepackage{amsthm,amsmath}
\usepackage[utf8]{inputenc} %unicode support
\usepackage{threeparttable}
\usepackage{enumerate}
\usepackage{url}
\usepackage[dvipdfmx]{graphicx}
\usepackage{color}

\startlocaldefs
\endlocaldefs

\begin{document}

%%% Start of article front matter
\begin{frontmatter}

\begin{fmbox}
\dochead{Full Paper}

%%%%%%%%%%%%%%%%%%%%%%%%%%%%%%%%%%%%%%%%%%%%%%
%%                                          %%
%% Enter the title of your article here     %%
%%                                          %%
%%%%%%%%%%%%%%%%%%%%%%%%%%%%%%%%%%%%%%%%%%%%%%

\title{Heliocentric Distance Dependence of Zodiacal Light Observed by Hayabusa2\#}

%%%%%%%%%%%%%%%%%%%%%%%%%%%%%%%%%%%%%%%%%%%%%%
%%                                          %%
%% Enter the authors here                   %%
%%                                          %%
%% Specify information, if available,       %%
%% in the form:                             %%
%%   <key>={<id1>,<id2>}                    %%
%%   <key>=                                 %%
%% Comment or delete the keys which are     %%
%% not used. Repeat \author command as much %%
%% as required.                             %%
%%                                          %%
%%%%%%%%%%%%%%%%%%%%%%%%%%%%%%%%%%%%%%%%%%%%%%

\author[
  addressref={aff1},                   % id's of addresses, e.g. {aff1,aff2}
  corref={aff1},                       % id of corresponding address, if any
% noteref={n1},                        % id's of article notes, if any
  email={ktsumura@tcu.ac.jp}   % email address
]{\inits{K.Ts.}\fnm{Kohji} \snm{Tsumura}}
\author[
  addressref={aff2},
  email={matsuura.shuji@kwansei.ac.jp}
]{\inits{S.M.}\fnm{Shuji} \snm{Matsuura}}
\author[
  addressref={aff3},
  email={sano.kei288@mail.kyutech.jp}
]{\inits{K.Sa.}\fnm{Kei} \snm{Sano}}
\author[
  addressref={aff4, aff5},
  email={iwata.takahiro@jaxa.jp}
]{\inits{T.I.}\fnm{Takahiro} \snm{Iwata}}
\author[
  addressref={aff4, aff5},
  email={yano.hajime@jaxa.jp}
]{\inits{H.Y.}\fnm{Hajime} \snm{Yano}}
\author[
  addressref={aff6},
  email={kitazato@u-aizu.ac.jp}
]{\inits{K.K.}\fnm{Kohei} \snm{Kitazato}}
\author[
  addressref={aff3},
  email={takimoto.koji670@mail.kyutech.jp}
]{\inits{K.Ta.}\fnm{Kohji} \snm{Takimoto}}
\author[
  addressref={aff7},
  email={manabu@perc.it-chiba.ac.jp}
]{\inits{M.Y.}\fnm{Manabu} \snm{Yamada}}
\author[
  addressref={aff8},
  email={morota@eps.s.u-tokyo.ac.jp}
]{\inits{T.M.}\fnm{Tomokatsu} \snm{Morota}}
\author[
  addressref={aff9},
  email={t.kouyama@aist.go.jp}
]{\inits{T.K.}\fnm{Toru} \snm{Kouyama}}
\author[
  addressref={aff4},
  email={hayakawa@planeta.sci.isas.jaxa.jp}
]{\inits{M.H.}\fnm{Masahiko} \snm{Hayakawa}}
\author[
  addressref={aff4},
  email={yokota@planeta.sci.isas.jaxa.jp}
]{\inits{Y.Y.}\fnm{Yasuhiro} \snm{Yokota}}
\author[
  addressref={aff10},
  email={etatsumi-ext@iac.es}
]{\inits{E.T.}\fnm{Eri} \snm{Tatsumi}}
\author[
  addressref={aff11},
  email={moe.matsuoka@aist.go.jp}
]{\inits{M.M.}\fnm{Moe} \snm{Matsuoka}}
\author[
  addressref={aff4},
  email={sakatani.naoya@jaxa.jp}
]{\inits{N.S.}\fnm{Naoya} \snm{Sakatani}}
\author[
  addressref={aff12},
  email={honda.rie.ha@ehime-u.ac.jp}
]{\inits{R.H.}\fnm{Rie} \snm{Honda}}
\author[
  addressref={aff13},
  email={kameda@rikkyo.ac.jp}
]{\inits{S.K.}\fnm{Shingo} \snm{Kameda}}
\author[
  addressref={aff14},
  email={suzuhide@meiji.ac.jp}
]{\inits{H.Su.}\fnm{Hidehiko} \snm{Suzuki}}
\author[
  addressref={aff8},
  email={cho@eps.s.u-tokyo.ac.jp}
]{\inits{Y.C.}\fnm{Yuichiro} \snm{Cho}}
\author[
  addressref={aff15},
  email={kazuo.yoshioka@edu.k.u-tokyo.ac.jp}
]{\inits{K.Y.}\fnm{Kazuo} \snm{Yoshioka}}
\author[
  addressref={aff16},
  email={ogawa.kazunori@jaxa.jp}
]{\inits{K.O.}\fnm{Kazunori} \snm{Ogawa}}
\author[
  addressref={aff17},
  email={kei.shirai@penguin.kobe-u.ac.jp}
]{\inits{K.Sh.}\fnm{Kei} \snm{Shirai}}
\author[
  addressref={aff4},
  email={sawada.hirotaka@jaxa.jp}
]{\inits{H.Sa.}\fnm{Hirotaka} \snm{Sawada}}
\author[
  addressref={aff8, aff7},
  email={sugita@eps.s.u-tokyo.ac.jp}
]{\inits{S.S.}\fnm{Seiji} \snm{Sugita}}
%%%%%%%%%%%%%%%%%%%%%%%%%%%%%%%%%%%%%%%%%%%%%%
%%                                          %%
%% Enter the authors' addresses here        %%
%%                                          %%
%% Repeat \address commands as much as      %%
%% required.                                %%
%%                                          %%
%%%%%%%%%%%%%%%%%%%%%%%%%%%%%%%%%%%%%%%%%%%%%%

\address[id=aff1]{%                           % unique id
  \orgdiv{Department of Natural Sciences, Faculty of Science and Engineering},             % department, if any
  \orgname{Tokyo City University},          % university, etc
  \postcode{158-8557}
  \city{Tokyo},                              % city
  \cny{Japan}                                    % country
}
\address[id=aff2]{%
  \orgdiv{Department of Physics, School of Science and Technology},
  \orgname{Kwansei Gakuin University},
  \postcode{669-1337}
  \city{Hyogo},
  \cny{Japan}
}
\address[id=aff3]{%
  \orgdiv{Department of Space Systems Engineering, School of Engineering},
  \orgname{Kyushu Institute of Technology},
  \postcode{804-8550}
  \city{Fukuoka},
  \cny{Japan}
}
\address[id=aff4]{%
  \orgdiv{Institute of Space and Astronautical Science},
  \orgname{Japan Aerospace Exploration Agency},
  \postcode{252-5210}
  \city{Kanagawa},
  \cny{Japan}
}
\address[id=aff5]{%
  \orgdiv{Department of Space and Astronautical Science},
  \orgname{The Graduate University of Advanced Studies},
  \postcode{240-0193}
  \city{Kanagawa},
  \cny{Japan}
}
\address[id=aff6]{%
  \orgdiv{Department of Computer Science and Engineering},
  \orgname{The University of Aizu},
  \postcode{965-8580}
  \city{Fukushima},
  \cny{Japan}
}
\address[id=aff7]{%
  \orgdiv{Planetary Exploration Research Center},
  \orgname{Chiba Institute of Technology},
  \postcode{275-0016}
  \city{Chiba},
  \cny{Japan}
}
\address[id=aff8]{%
  \orgdiv{Department of Earth and Planetary Science},
  \orgname{University of Tokyo},
  \postcode{113-0033}
  \city{Tokyo},
  \cny{Japan}
}
\address[id=aff9]{%
  \orgdiv{Digital Architecture Research Center},
  \orgname{National Institute of Advanced Industrial Science and Technology},
  \postcode{135-0064}
  \city{Tokyo},
  \cny{Japan}
}
\address[id=aff10]{%
  \orgdiv{nstituto de Astrofisica de Canarias},
  \orgname{University of La Laguna},
  \city{Tenerife},
  \cny{Spain}
}
\address[id=aff11]{%
  \orgdiv{Geological Survey of Japan},
  \orgname{National Institute of Advanced Industrial Science and Technology},
  \postcode{305-8567}
  \city{Ibaraki},
  \cny{Japan}
}
\address[id=aff12]{%
  \orgdiv{Center for Data Science},
  \orgname{Ehime University},
  \postcode{790-8577}
  \city{Ehime},
  \cny{Japan}
}
\address[id=aff13]{%
  \orgname{Rikkyo University},
  \postcode{171-8501}
  \city{Tokyo},
  \cny{Japan}
}
\address[id=aff14]{%
  \orgname{Meiji University},
  \postcode{214-8571}
  \city{Kanagawa},
  \cny{Japan}
}
\address[id=aff15]{%
  \orgdiv{Department of Complexity Science and Engineering},
  \orgname{University of Tokyo},
  \postcode{277-8561}
  \city{Chiba},
  \cny{Japan}
}
\address[id=aff16]{%
  \orgdiv{JAXA Space Exploration Center},
  \orgname{Japan Aerospace Exploration Agency},
  \postcode{252-5210}
  \city{Kanagawa},
  \cny{Japan}
}
\address[id=aff17]{%
  \orgname{Kobe University},
  \postcode{657-8501}
  \city{Hyogo},
  \cny{Japan}
}
%%%%%%%%%%%%%%%%%%%%%%%%%%%%%%%%%%%%%%%%%%%%%%
%%                                          %%
%% Enter short notes here                   %%
%%                                          %%
%% Short notes will be after addresses      %%
%% on first page.                           %%
%%                                          %%
%%%%%%%%%%%%%%%%%%%%%%%%%%%%%%%%%%%%%%%%%%%%%%

%\begin{artnotes}
%%\note{Sample of title note}     % note to the article
%\note[id=n1]{Equal contributor} % note, connected to author
%\end{artnotes}

%\end{fmbox}% comment this for two column layout

%%%%%%%%%%%%%%%%%%%%%%%%%%%%%%%%%%%%%%%%%%%%%%%
%%                                           %%
%% The Abstract begins here                  %%
%%                                           %%
%% Please refer to the Instructions for      %%
%% authors on https://www.biomedcentral.com/ %%
%% and include the section headings          %%
%% accordingly for your article type.        %%
%%                                           %%
%%%%%%%%%%%%%%%%%%%%%%%%%%%%%%%%%%%%%%%%%%%%%%%

\begin{abstractbox}

\begin{abstract} % abstract
Zodiacal light (ZL) is sunlight scattered by interplanetary dust particles (IDPs) at optical wavelengths.
The spatial distribution of IDPs in the Solar System may hold an important key 
to understanding the evolution of the Solar System and material transportation within it.
The number density of IDPs can be expressed as $n(r) \sim r^{-\alpha}$,
and the exponent $\alpha \sim 1.3$ was obtained by previous observations from interplanetary space by Helios 1/2 and Pioneer 10/11 in the 1970s and 1980s.
However, no direct measurements of $\alpha$ based on ZL observations from interplanetary space outside Earth's orbit have been performed since then. 
Here, we introduce initial results for the radial profile of the ZL at optical wavelengths observed over the range 0.76-1.06~au by ONC-T aboard the Hayabusa2\# mission in 2021-2022.
The ZL brightness we obtained is well reproduced by a model brightness, although there is a small excess of the observed ZL brightness over the model brightness at around 0.9~au.
The radial power-law index we obtained is $\alpha = 1.30 \pm 0.08$, which is consistent with previous results based on ZL observations.
The dominant source of uncertainty arises from the uncertainty in estimating the diffuse Galactic light (DGL).
\end{abstract}

%%%%%%%%%%%%%%%%%%%%%%%%%%%%%%%%%%%%%%%%%%%%%%
%%                                          %%
%% The keywords begin here                  %%
%%                                          %%
%% Put each keyword in separate \kwd{}.     %%
%%                                          %%
%%%%%%%%%%%%%%%%%%%%%%%%%%%%%%%%%%%%%%%%%%%%%%


% MSC classifications codes, if any
%\begin{keyword}[class=AMS]
%\kwd[Primary ]{}
%\kwd{}
%\kwd[; secondary ]{}
%\end{keyword}

\end{abstractbox}
\end{fmbox}% uncomment this for two column layout

\end{frontmatter}

%%%%%%%%%%%%%%%%%%%%%%%%%%%%%%%%%%%%%%%%%%%%%%%%
%%                                            %%
%% The Main Body begins here                  %%
%%                                            %%
%% Please refer to the instructions for       %%
%% authors on:                                %%
%% https://www.biomedcentral.com/getpublished %%
%% and include the section headings           %%
%% accordingly for your article type.         %%
%%                                            %%
%% See the Results and Discussion section     %%
%% for details on how to create sub-sections  %%
%%                                            %%
%% use \cite{...} to cite references          %%
%%  \cite{koon} and                           %%
%%  \cite{oreg,khar,zvai,xjon,schn,pond}      %%
%%                                            %%
%%%%%%%%%%%%%%%%%%%%%%%%%%%%%%%%%%%%%%%%%%%%%%%%

%%%%%%%%%%%%%%%%%%%%%%%%% start of article main body
% <put your article body there>

\textbf{Graphical Abstract}
\begin{center}
\includegraphics[width=16.5cm]{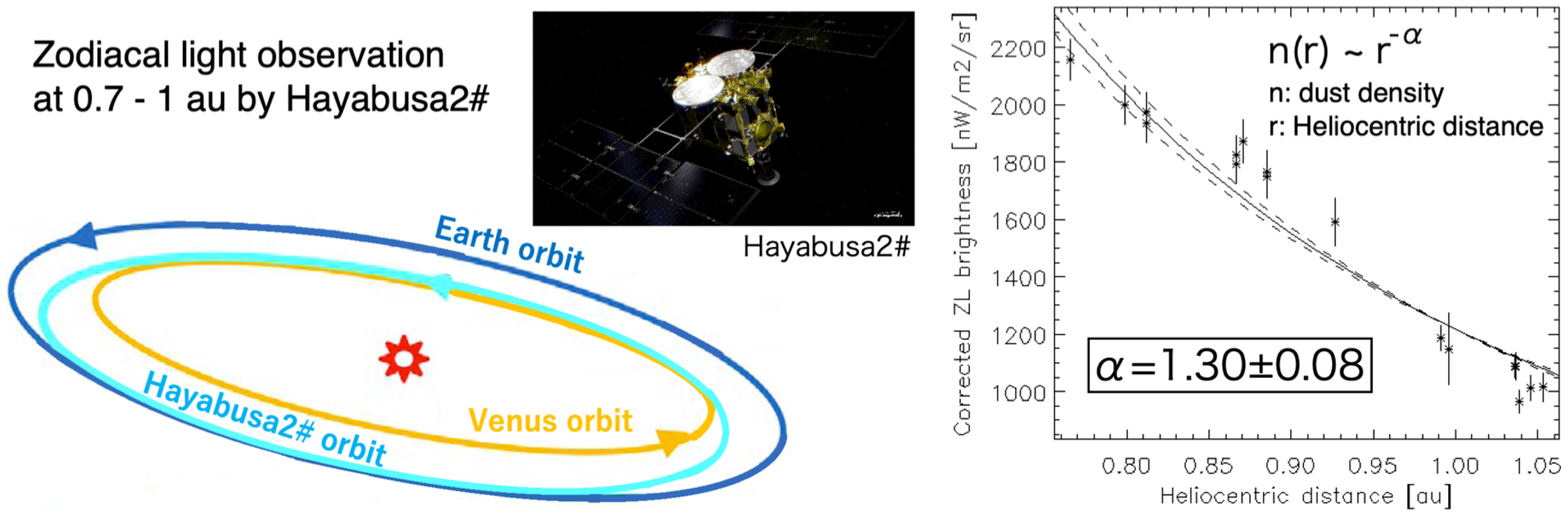} 
\end{center}
\clearpage

%%%%%%%%%%%%%%%%
%% Background %%
\section{Introduction}
The zodiacal light (ZL) is sunlight scattered by interplanetary dust particles (IDPs) at optical wavelengths, and it is a major constituent of the diffuse celestial brightness. 
A continuous supply of IDPs is necessary to sustain the diffuse brightness because IDP is removed from the Solar System due to the Poynting-Robertson (PR) effect and by radiation pressure from the Sun \cite{Wyatt1950, BURNS19791}.
Possible sources for this supply are asteroid collisions \cite{Dermott1984, Schramm1989, Tsumura2010} or cometary ejections \cite{LIOU1995717, Nesvorny2010, Yang2015}, 
but the relative ratios of the contributions from these sources are still unknown. 
Dust of interstellar origin also contributes $\sim$10\% to the total amount of IDP \cite{Rowan2013}.
Thus, observational constraints that can tell the differences among these sources are important for a better understanding the origin and characteristics of IDPs and of the way planetary and exoplanetary systems evolve with time \cite{Leinert1998, LASUE2020104973}.

Historically, extensive ZL observations were conducted from ground-based telescopes at high-altitude sites in the 1960s and 1970s \cite{Dumont1975, Levasseur1980},
but the accuracy of these ZL observations is limited due to atmospheric emission.
In contrast, space-based platforms eliminate atmospheric contamination and provide precise ZL measurements
\cite{Murdock1985, MATSUURA1995, Matsumoto1996, Tsumura2010, Tsumura2013a, BUFFINGTON201688, Korngut2022, Takimoto2022, Takimoto2023}.
The ZL is the only sky-brightness component that is not fixed on the celestial sphere.
In general, the ZL is smoothly distributed, and its small-scale spatial structures are only at the level of a few percent owing to the smooth spatial distribution of IDP as a smooth cloud \cite{Pyo2012}.
The plane of symmetry of the smooth cloud is slightly inclined to the ecliptic plane because of the Jovian orbit.
Seasonal variations in ZL occur for an Earth-based observer due to the orbital motion of the Earth, 
which changes the heliocentric distance and the position of the observer with respect to the symmetry plane.
A detailed IDP distribution model has been established based on the seasonal variation of the ZL \cite{Kelsall1998, Wright1998}.

The number density ($n$) of IDP is presumed to be of a form that is separable into radial and vertical terms;
\begin{equation} n(r, \beta) = n_{0}\left(\frac{r}{r_0}\right)^{-\alpha } f(\beta), \end{equation}
where $n_0$ is the reference number density of IDPs in the symmetry plane at the heliocentric distance $r_0$, 
and $f(\beta)$ denotes the vertical distribution as a function of an elevation angle $\beta$ from the symmetry plane \cite{GIESE1986395}.
The assumption that the vertical distribution of the IDPs depends only on $\beta$ is suggested by the fact that the PR effect does not affect the orbital inclinations of particles as they spiral into the Sun.
The radial power-law is induced by the radial distribution expected for particles under the influence of the PR effect, which results in $\alpha =1$ for dust bound in a circular orbit \cite{BURNS19791}.
When dust-grain sizes are reduced by sublimation near the Sun,
such smaller dust particles are expelled from the Solar System as $\beta$-meteoroids by radiation pressure \cite{ZOOK1975183, Wehry1999, KRUGER2014657}. 
The radial profile of $\beta$-meteoroids is expected to follow a power law with $\alpha =2$ \cite{Szalay2020}.
The relative ratio of these two components remains an open issue and may hold an important key for understanding the evolution of the IDP distribution \cite{Leinert1990, Mann2004}.

The ZL brightness $I_{ZL}$ can be modeled as the integral of scattered sunlight along the line of sight:
\begin{equation} I_{ZL} = \int F_{\odot}(r)  n(r) A \Phi(\theta) dl, \end{equation}
where $F_{\odot}(r) \sim r^{-2}$ is the Solar flux at the distance $r$ from the Sun, $A$ is the albedo of the IDP, $\Phi(\theta)$ is the phase function at the scattering angle $\theta$,
and $dl$ is an increment along the line of sight.
If the scattering properties (size and albedo) of IDPs do not change significantly with heliocentric distance, 
the heliocentric dependence of ZL toward the antisolar direction on the symmetry plane can be written as $I_{ZL} \sim r^{-(\alpha +1)}$.
If the line of sight is not oriented in the antisolar direction, the heliocentric dependence of $I_{ZL}$ becomes much more complex, since it depends on the phase function $\Phi(\theta)$ for which $\theta$ will vary.
In addition, a heliocentric dependence of the local albedo of the IDPs has also been reported \cite{Levasseur1991}, which makes the heliocentric dependence of $I_{ZL}$ even more complex.

Direct observations of the radial power-law index $\alpha$  based on ZL observations were performed from spacecraft outside Earth's orbit in the 1970s and 1980s. 
Pioneer 10/11 observations of the ZL at 1-3.3~au gave $\alpha =$1-1.5.
More specifically, a single power-law model with $\alpha \sim 1$ and a cutoff near 3.3~au gives the best fit to the observational data,
although a two-component model with $\alpha \sim 1.5$ and increased IDP in the asteroid belt fits the data equally well \cite{Hanner1976}.
Helios 1/2 observations of the ZL at 0.3-1~au gave $\alpha =1.3 \pm 0.05$,
although $\alpha = 1.35$ gives a better fit for small solar elongations ($<50 ^{\circ}$), and $\alpha = 1.25$ is more appropriate for large solar elongations ($>100 ^{\circ}$) \cite{Leinert1981, Leinert1982}.
ZL observations from spacecraft outside Earth's orbit have not been performed following these missions. 
The Japanese Venus orbiter Akatsuki tried but could not detect the ZL due to insufficient cooling of the sensor \cite{Satoh2016}.

Some IDP distribution models were developed based on observations of the all-sky ZL brightness and its seasonal variation from geocentric orbit.
In particular, observations from the Cosmic Background Explorer (COBE) yielded $\alpha = 1.34 \pm 0.022$ \cite{Kelsall1998} and $\alpha = 1.22$ \cite{Wright1998}, and observations by AKARI gave $\alpha = 1.59 \pm 0.02$ \cite{Kondo2016}. 
These observations of the ZL were performed at 1~au, so the accuracy in determining $\alpha$ was worse than that obtained by direct observations from interplanetary space.

The value of $\alpha$ has also been determined based on the observations of the inner ZL or F-corona.
Observations of the inner ZL by Clementine from lunar orbit while the Sun was in eclipse behind the Moon yielded $\alpha = 1.45 \pm 0.05$  \cite{HAHN2002360}.
Values of $\alpha$ from 1.31 to 1.35 were obtained from F-corona observations at elongations ranging from 0.07 to 0.45 au from the Sun \cite{Stenborg2018}
by the Heliospheric Imager-1 \cite{Eyles2009} onboard the Solar TErrestrial RElations Observatory-A (STEREO-A) orbiting the Sun at approximately 1~au.
In addition, $\alpha = 1.31$ was obtained by F-corona observations between 0.1 and 0.4 au \cite{Howard2019}
by the Widefield Imager for Solar Probe inner telescope (WISPER-1) \cite{Vourlidas2016} onboard the Parker Solar Probe (PSP) when it passed perihelion at 0.16-0.25 au. 
These results are limited to dust distributions close to the Sun.

A technique for studying the distribution and properties of IDPs independent of the ZL observation is in-situ dust counting using dedicated dust detectors.
The size distribution of IDPs was studied by the in-situ dust-counting method and it was suggested that large (10-100~$\mu$m) dust is dominant around 1~au \cite{GRUN1985244, Divine1993}.
The ZL brightness is indicative of the IDP distribution in the inner Solar System, where the IDP density is substantial, and the IDP distribution derived from these ZL observations is confined to the inner Solar System ($<5$~au).
Conversely, dust distribution in the outer Solar System has been investigated by the in-situ dust-counting method \cite{Poppe2019, Bernardoni2022}.

This paper introduces the IDP distribution based on the ZL observations from the Hayabusa2\# mission at 0.76-1.06~au performed in 2021-2022.
These are the first successful observations of the ZL from outside Earth's orbit in the last 40 years.

\begin{figure}
\begin{center}
\includegraphics[width=8cm]{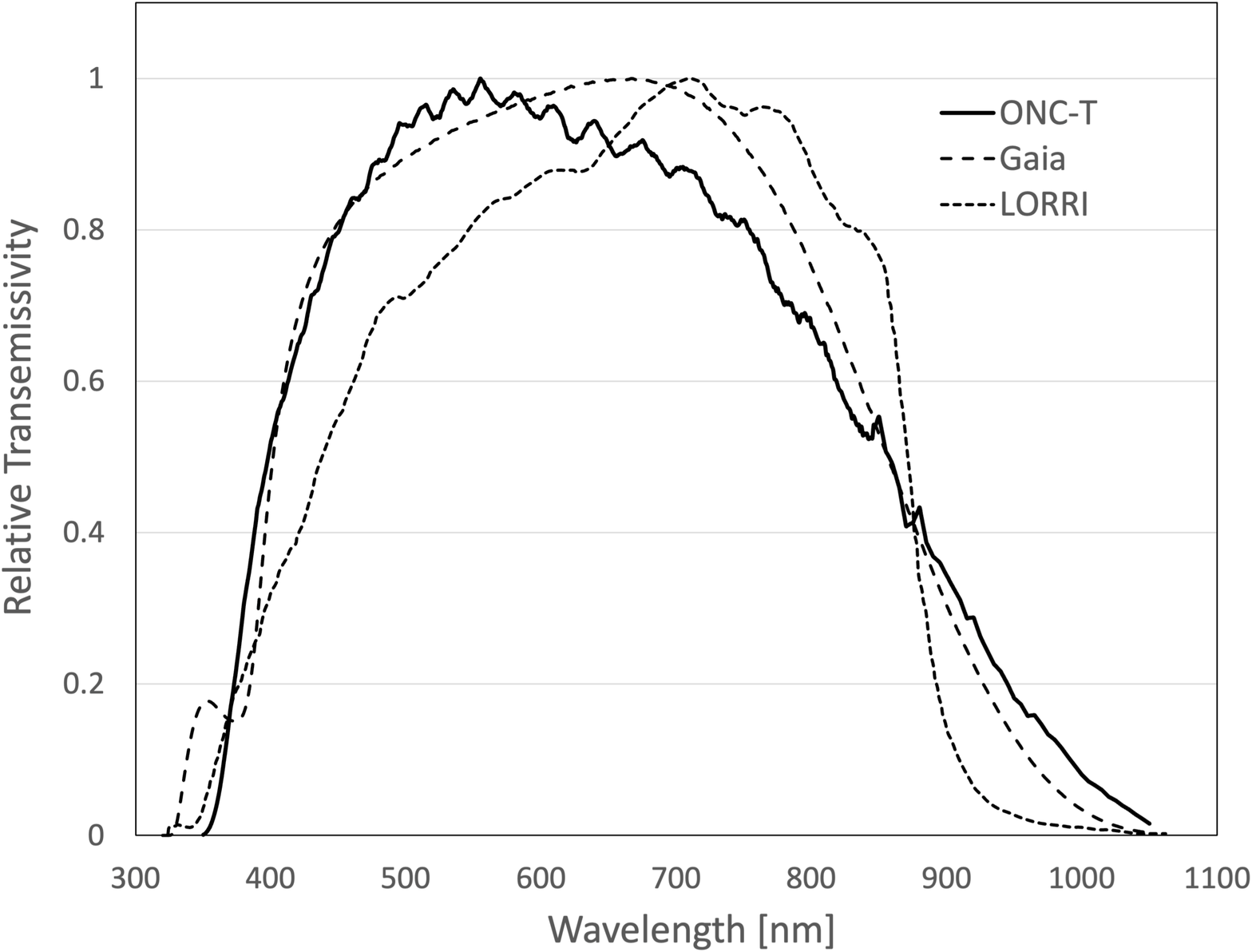} 
\end{center}
\caption{A comparison of the relative transmissivity. The solid line shows the bandpass of ONC-T/Hayabusa2\# wide-band \cite{TATSUMI2019153},
the dashed line shows the bandpass of Gaia G-band \cite{Riello}, and the dotted line shows the bandpass of LORRI/New Horizons \cite{Cheng2008}.}
\label{filter}
\end{figure}

\section{Data Acquisition and Reduction}
\subsection{Hayabusa2\# overview}
Hayabusa2 is the second Japanese asteroid-sample-return mission. 
The Hayabusa2 spacecraft was launched in December 2014 and successfully arrived at asteroid (162173) Ryugu in June 2018.
After extensive scientific observations for $\sim$1.5 years, it departed from Ryugu in November 2019 
and successfully brought the capsule containing Ryugu samples back to Earth in December 2020 \cite{TSUDA20225, Tachibana2022}. 
With the successful main mission of the sample return completed,
an extended mission named Hayabusa2\# (SHARP; Small Hazardous Asteroid Reconnaissance Probe) was initiated to explore new asteroids;
it will perform a fly-by of (98943) 2001 CC21 in July 2026 and a rendezvous with 1998 KY26 in July 2031 \cite{MIMASU2022557}.
Some scientific observations including ZL observations will be performed during this long cruising phase \cite{HIRABAYASHI20211533}.

The Optical Navigation Camera (ONC) onboard Hayabusa2 consists of one telescopic camera (ONC-T) and two wide-angle view cameras (ONC-W1/W2) \cite{Kameda2017, SUZUKI2018341, TATSUMI2019153, KOUYAMA2021114353, Yamada2023},
and it was used for both global and local high-resolution optical observations of Ryugu \cite{Sugita2019}.
The ONC was carefully calibrated both before and after launch, and it remains in good condition after contact with the surface of Ryugu during the two touchdowns for sampling.
In this study, we used ONC-T for the ZL observations.
The field of view of ONC-T is $6.27 \times 6.27$~deg$^2$, which is covered with a $1024 \times 1024$~pixel region of a CCD detector \cite{Kameda2017}.
The longest exposure time of ONC-T was 178~sec, which we used for the ZL observations in this study.
ONC-T has a wheel system that rotates seven color-bandpass filters and one wide clear filter (a panchromatic glass window).
We used the wide-band filter ($\lambda$ = 612~nm and $\Delta \lambda$ = 448~nm, see Figure~\ref{filter}) for the ZL observations,
with a v-band filter ($\lambda$ = 550~nm and $\Delta \lambda$ = 28~nm) for stray light subtraction (see Section \ref{sec:stray}).

\begin{figure}
\begin{center}
\includegraphics[width=8cm]{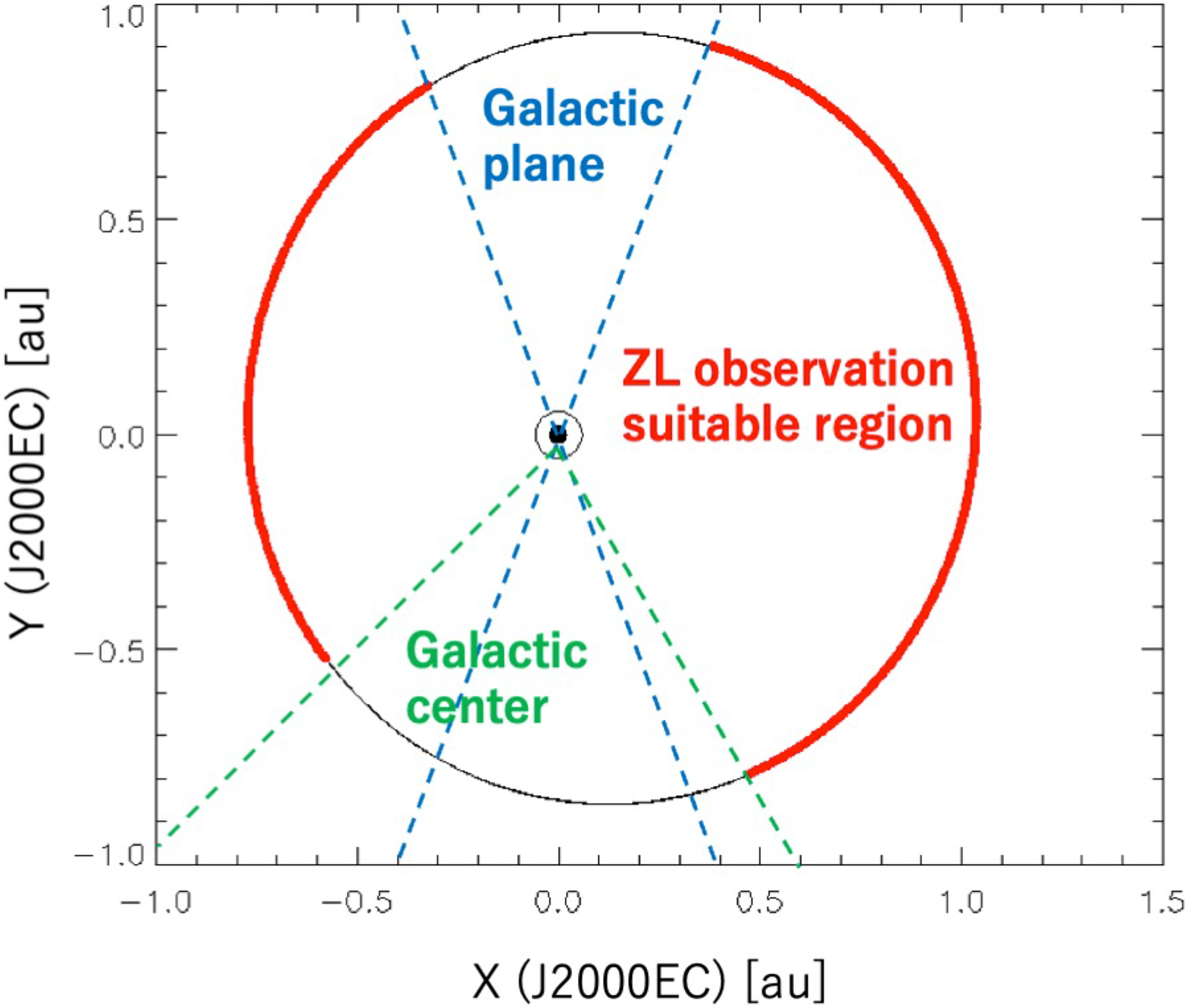} 
\end{center}
\caption{Hayabusa2\# orbit in the J2000EC inertial frame before the Earth swing-by on December 2027. 
The regions suitable for ZL observations are shown in red,
and unsuitable regions owing to pointing toward the Galactic plane (Galactic latitude $< 20^{\circ}$) and the Galactic center (Galactic longitude $< 20^{\circ}$) are shown in blue and green, respectively.}
\label{orbit}
\end{figure}

\begin{figure*}
\begin{center}
\includegraphics[width=16.5cm]{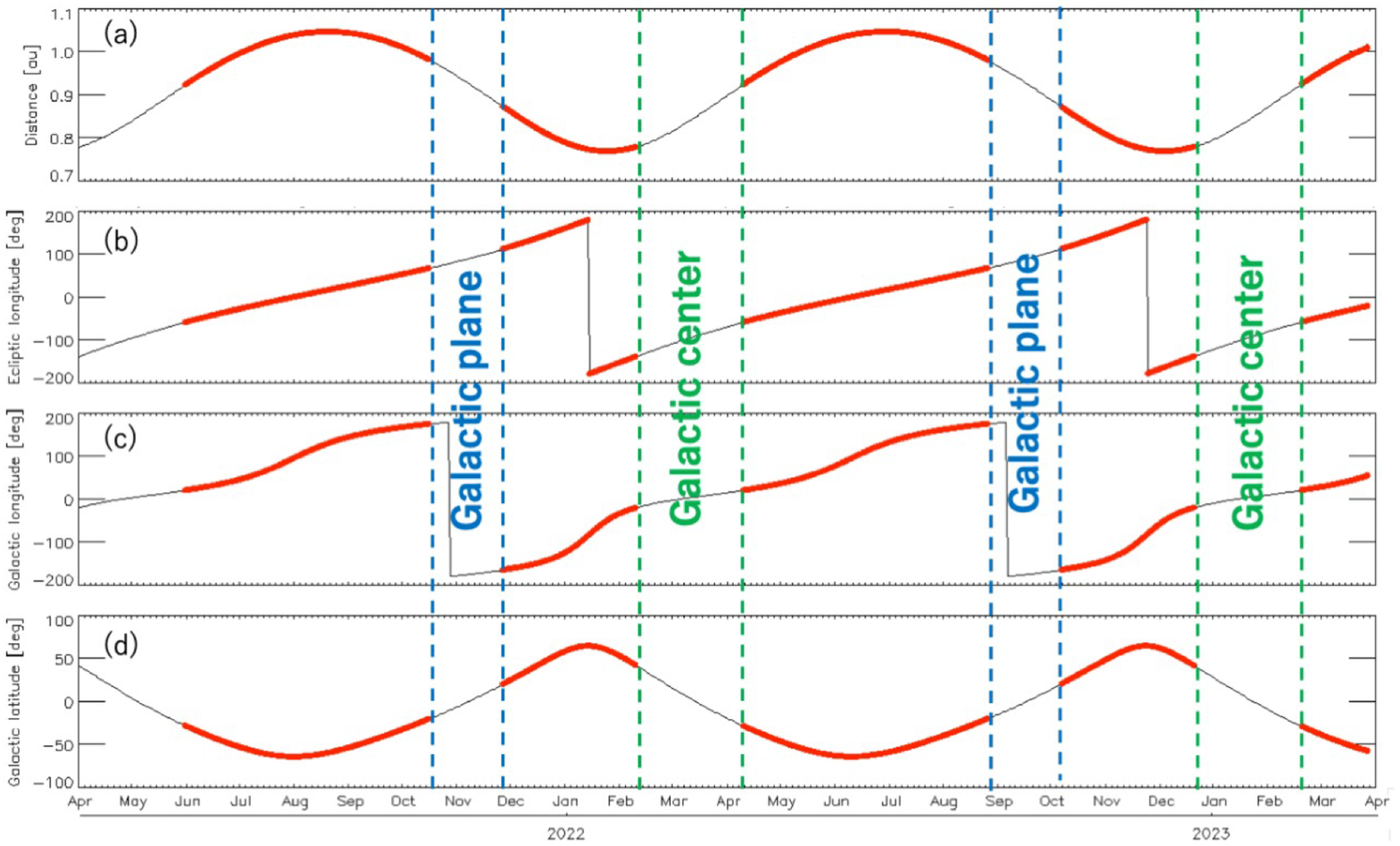} 
\end{center}
\caption{(a) Heliocentric distance of the Hayabusa2 spacecraft and (b) its ecliptic longitude, (c) Galactic longitude, and (d) Galactic latitude of the antisolar direction during the period from April 2021 to March 2023. 
The regions suitable for ZL observations are shown in red, 
and unsuitable periods owing to pointing toward the Galactic plane (Galactic latitude $< 20^{\circ}$) and the Galactic center (Galactic longitude $< 20^{\circ}$) are shown in blue and green, respectively. }
\label{orbit2}
\end{figure*}

\begin{table*}
\renewcommand{\arraystretch}{1.2}
\begin{center}
\vskip 1cm
\caption{Spacecraft positions and observed fields}
\begin{tabular}{l|cccc|cccc} 
  \hline\noalign{\vskip3pt} 
   & \multicolumn{4}{c|}{Position of Spacecraft [au]} & \multicolumn{4}{c}{Observed Field [deg]} \\  
   Date & X & Y & Z & R$^{(1)}$ & (RA, Dec) & (Elon, Elat)$^{(2)}$ & (Glon, Glat)$^{(3)}$ & Solar Elongation\\
   \hline\noalign{\vskip3pt} 
2021-08-23 &  1.023 &  0.252 & -0.069 & 1.053 &  ( 15.93,  -3.65) & ( 13.25,  -9.64) & (130.58, -66.34) & 174.10 \\
2021-09-20 &  0.821 &  0.633 & -0.047 & 1.037 &  ( 37.06,   5.46) & ( 36.52,  -8.71) & (162.23, -49.88) & 173.81 \\
2021-11-29 & -0.261 &  0.846 &  0.035 & 0.886 &  (104.42,  18.72) & (103.67,  -4.03) & (196.79,   9.75) & 172.78 \\
2021-12-06 & -0.375 &  0.781 &  0.042 & 0.866 &  (112.96,  18.16) & (111.80,  -3.56) & (200.79,  16.88) & 172.57 \\
2021-12-28 & -0.664 &  0.468 &  0.057 & 0.812 &  (142.74,  12.54) & (141.02,  -2.05) & (219.94,  41.00) & 172.84 \\
2022-01-24$^{(4)}$ & -0.774 & -0.067 &  0.055 & 0.777 &  (198.38,   3.56) & (195.57,  10.49) & (316.49,  65.85) & 167.64 \\
2022-02-14$^{(4)}$ & -0.625 & -0.474 &  0.037 & 0.784 &  (204.83, -14.26) & (208.19,  -3.68) & (320.08,  47.01) & 169.01 \\
2022-04-18 &  0.451 & -0.810 & -0.048 & 0.927 &  (297.25, -31.48) & (293.38, -10.22) & (  8.92, -25.26) & 170.79 \\
2022-05-16 &  0.834 & -0.535 & -0.071 & 0.991 &  (327.89, -25.02) & (321.52, -11.33) & ( 25.45, -49.77) & 170.76 \\
2022-06-20 &  1.039 & -0.024 & -0.076 & 1.039 &  (357.99, -13.31) & (352.80, -11.40) & ( 74.84, -70.40) & 170.74 \\
2022-07-04 &  1.028 &  0.193 & -0.071 & 1.046 &  (  8.76,  -8.29) & (  4.74, -11.08) & (110.55, -70.76) & 170.74 \\
2022-08-01 &  0.854 &  0.587 & -0.051 & 1.036 &  ( 30.13,   1.62) & ( 28.62, -10.00) & (155.74, -56.78) & 170.76 \\
2022-08-29 &  0.505 &  0.858 & -0.021 & 0.996 &  ( 53.46,  10.57) & ( 53.73,  -8.38) & (174.72, -35.49) & 170.80 \\
2022-10-17 & -0.318 &  0.811 &  0.038 & 0.871 &  (106.56,  17.81) & (105.80,  -4.72) & (198.51,  11.20) & 170.88 \\
2022-11-14 & -0.683 &  0.415 &  0.057 & 0.799 &  (144.53,  10.87) & (143.23,  -3.08) & (222.98,  41.84) & 170.95 \\
2022-12-12 & -0.750 & -0.154 &  0.052 & 0.766 &  (184.25,  -5.34) & (186.02,  -3.21) & (287.28,  56.48) & 171.00 \\
  \hline\noalign{\vskip3pt}
  \end{tabular}
  \begin{tablenotes}
\item[1] (1) $R = \sqrt{X^2 +Y^2}$.
\item[2] (2) Ecliptic coordinate.
\item[3] (3) Galactic coordinate.
\item[4] (4) These data were excluded from the analysis owing to the presence of additional stray light (see Section \ref{sec:stray}).
\end{tablenotes}
\label{tab:ObsField}
\end{center}
\end{table*}

\subsection{Observation Fields}\label{sec:field}
The Hayabusa2 spacecraft followed an elliptical orbit over the range 0.76-1.06~au before the Earth swing-by on December 2027, as shown in Figure~\ref{orbit} and \ref{orbit2} \cite{MIMASU2022557}.
The spacecraft maintained an attitude in which the solar-array paddle (+Z direction) was pointed toward the Sun during this period, 
and we performed the ZL observations during periods when the ion engines were not in operation.
Since ONC-T points toward the -Z direction, the ZL is observed toward the antisolar direction. 
This is an advantage of our ZL observations over past observations because previous ZL observations in interplanetary space were made at various solar elongation angles, 
making it difficult to distinguish whether the ZL changes were due to changes in the heliocentric distance or in the solar elongation.
In our ZL observations, the change in ZL brightness due to the solar elongation was minimized by observing the ZL at a nearly constant solar elongation (see Table \ref{tab:ObsField}).

Stray light is produced when sunlight hits the radiator that cools the ONC-T detector from a certain range of directions \cite{SUZUKI2018341, TATSUMI2019153}.
Thus, the ZL observations need to be conducted in a ``stray-light-avoidance attitude'', in which the -X side and +Y side of the spacecraft are illuminated by the Sun. 
For this reason, the actual directions of our ZL observations are shifted from the antisolar direction by $\sim$10 degrees.
Table \ref{tab:ObsField} summarizes the observed fields and the position of the spacecraft when the observations were conducted.

Periods when the observable direction (antisolar direction) is pointed toward either the Galactic plane or the Galactic center are not suitable for ZL observations because the Galactic brightness is too strong (see Section \ref{sec:ISL} and \ref{sec:DGL}).
Thus, we define as unsuitable ZL observation periods those in which the antisolar direction is pointing toward (1) Galactic latitude $< 20^{\circ}$ (the blue zones in Figures~\ref{orbit} and \ref{orbit2})
or (2) Galactic longitude $< 20^{\circ}$ (the green zones in Figures~\ref{orbit} and \ref{orbit2}).
The ZL observations were made approximately once a month during the time suitable for ZL observations (the red zones in Figures~\ref{orbit} and \ref{orbit2}).

\subsection{Acquired Images} \label{sec:data}
The ONC-T detector has a $1024 \times 1024$ pixel imaging region, with a $16 \times 1024$ pixel masked regions termed ``optical black'' on each side as a dark reference \cite{Kameda2017}. 
The two optical-black images are combined and treated as one $32 \times 1024$ pixel optical-black image.
Raw images acquired by ONC-T are processed in a sequence of steps to calibrate the image data.
In this work, we used L2a-level images, which are raw FITS images with header information containing the spacecraft system housekeeping data and ONC status data 
(the temperatures of the detector, lens system, and electronics as well as the voltages of the electronics, etc.) \cite{TATSUMI2019153}.
The signal from each pixel is provided in 16-bit digital numbers (DN).

\begin{figure}
\begin{center}
\includegraphics[width=7.5cm]{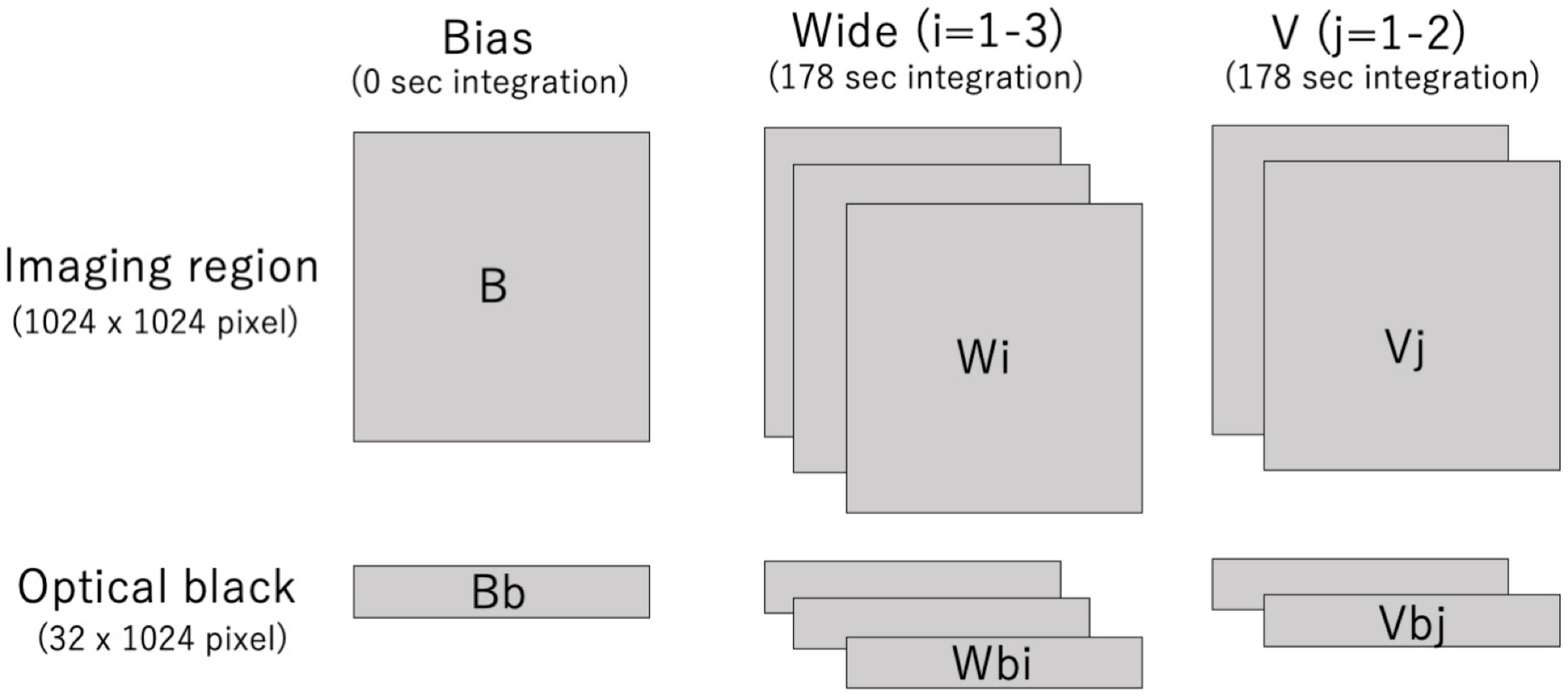} 
\end{center}
\caption{Dataset acquired for one ZL observation. One dataset includes one bias image $B(x,y)$, three wide-band images $W_i(x,y)$ ($i=$1-3), 
two v-band images $V_j(x,y)$ ($j=$1-2), and their respective optical-black images ($Bb(x,y)$, $Wb_i(x,y)$, and $Vb_j(x,y)$).}
\label{images}
\end{figure}

One ZL observation dataset includes one bias image $B(x,y)$, three wide-band images $W_i(x,y)$ ($i=$1-3), two v-band images $V_j(x,y)$ ($j=$1-2), 
and their respective optical-black images ($Bb(x,y)$, $Wb_i(x,y)$, and $Vb_j(x,y)$), as shown in Figure~\ref{images}.
For the ZL observations on 2021-11-29, 2021-12-06, 2021-12-28, 2022-01-24, and 2022-02-14, as part of the calibration operations we acquired two data sets to monitor the stability of the ONC-T sensitivity after it was turned on.

\subsection{Dark-Current Subtraction}  \label{sec:dark}
\begin{figure}
\begin{center}
\includegraphics[width=8cm]{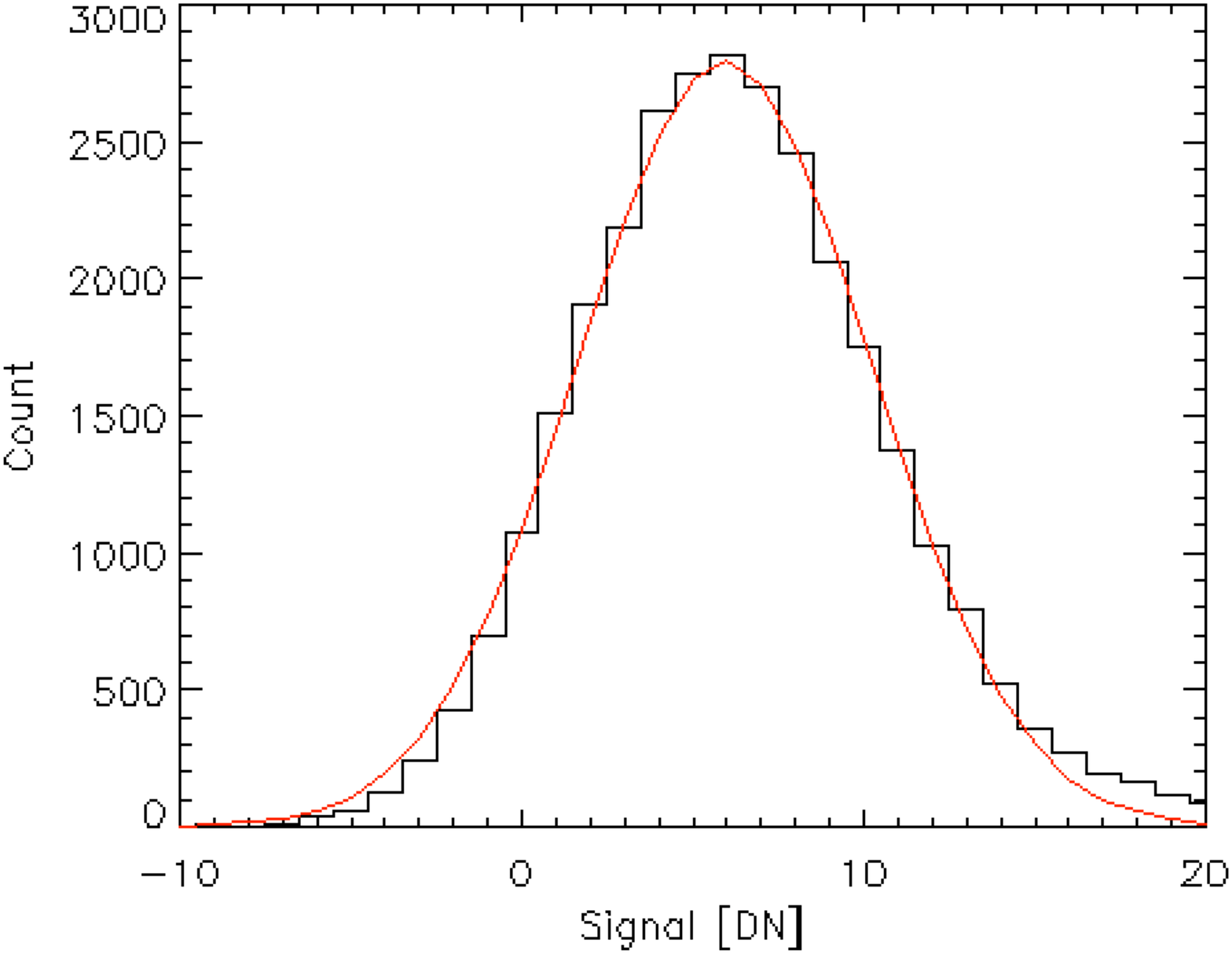} 
\end{center}
\caption{Dark-current estimation. A histogram of the first bias-subtracted optical-black image, $Wb_1(x,y)-Bb(x,y)$, obtained on 2022-08-29 (black) and its best-fit Gaussian function (red). 
We obtained the dark current as $I_{\textrm{dark}^{W_1}}=5.92$~DN from the peak position of this histogram.}
\label{dark}
\end{figure}

Dark-current subtraction is essential for the measurement of diffuse radiation such as the ZL.
We estimated the dark current for each image in our dataset from the corresponding dark image, 
which we obtained from the optical-black image by subtracting a bias image ($Wb_i(x,y)-Bb(x,y)$ and $Vb_j(x,y)-Bb(x,y)$).
We then created a histogram of the dark image, and we take its peak position to be the value of the dark current for that image. 
We fitted the histogram of the dark image with a Gaussian function.
The peak position of the histogram corresponds to the mode of the dark image.
Using this Gaussian-fitting procedure, we eliminated bad pixels such as those due to leakage of light from the imaging region or due to hot pixels caused by cosmic-ray hits. 
We expressed the resulting dark-current values for the wide-band and v-band images as $I_{\textrm{dark}}^{W_i}$ and $I_{\textrm{dark}}^{V_j}$, respectively.
As an example, we found the dark current of the first bias-subtracted optical-black image taken on 2022-08-29 to be $I_{\textrm{dark}}^{W_1} = 5.92$~DN from the peak position of the histogram, as shown in Figure~\ref{dark}.

Next, the obtained dark current and the bias image are subtracted from the imaging region, 
yielding three wide-band subtracted images and two v-band subtracted images in one data set. 
From the wide-band images, we generated a single reduced image $W(x,y$) from three subtracted images by taking the median of each pixel:
\begin{equation} W(x,y) = \textrm{median}[W_i(x,y) - (B(x,y) + I_{\textrm{dark}}^{W_i})]. \end{equation}
This median procedure removes many hot pixels caused by cosmic-ray hits.
Since there are only two v-band images, we cannot use the median procedure for them.
Instead, we generated a single reduced image $V(x,y)$ by taking the minimum of each pixel to reduce hot pixels caused by cosmic-ray hits:
\begin{equation} V(x,y) = \textrm{min}[V_j(x,y) - (B(x,y) + I_{\textrm{dark}}^{V_j})].\end{equation}

\subsection{Stray-Light subtraction} \label{sec:stray}
\begin{figure*}
\begin{center}
\includegraphics[width=16.3cm]{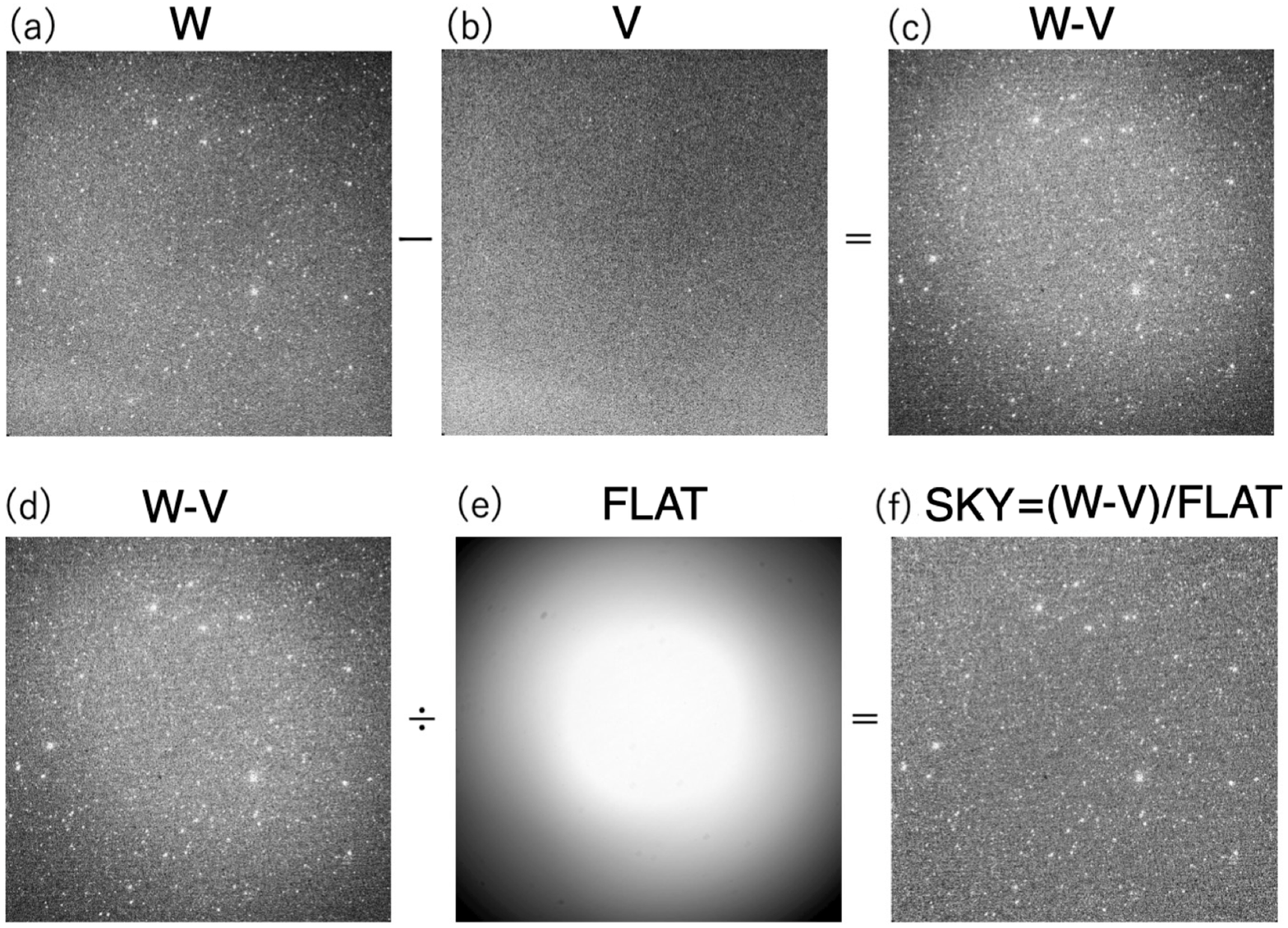} 
\end{center}
\caption{Data-reduction procedure. (a): The wide-band image generated by taking the median of the three dark-subtracted wide-band images. 
(b): The v-band image generated by taking the minimum of two dark-subtracted v-band images as a reference frame for the stray light.
(c): The stray-light-subtracted image $W(x,y)-V(x,y)$. (d): Same as (c). (e): A normalized flat-field image \cite{SUZUKI2018341}.
(f): The flat-field-corrected image $SKY(x,y)$, which is the final reduced image of the sky used for scientific analysis.}
\label{stray}
\end{figure*}

Since stray light occurs in ONC-T images when the spacecraft is at certain attitudes \cite{SUZUKI2018341, TATSUMI2019153}, 
we performed all ZL observations in stray-light-avoidance attitudes (Section \ref{sec:field}).
However, weak stray light remains even in this case.
Figures~\ref{stray}a and \ref{stray}b the reduced wide-band image $W(x,y)$ and v-band image $V(x,y)$ obtained on 2021-08-23, respectively,
and the stray-light patterns can be seen clearly in these images.
It is known that the intensity and pattern of the stray light do not depend on the filter selection \cite{SUZUKI2018341}.
Thus, we subtracted the v-band image $V(x,y)$ as a stray-light reference frame from the wide-band image $W(x,y)$ to remove the remaining stray light, as shown in Figures~\ref{stray}c and \ref{stray}d.
Since the ZL signal in the v-band image is estimated to be less than 1~DN, there is little impact on the scientific analysis of the ZL due to this stray-light-removal procedure.

\begin{figure}
\begin{center}
\includegraphics[width=6cm]{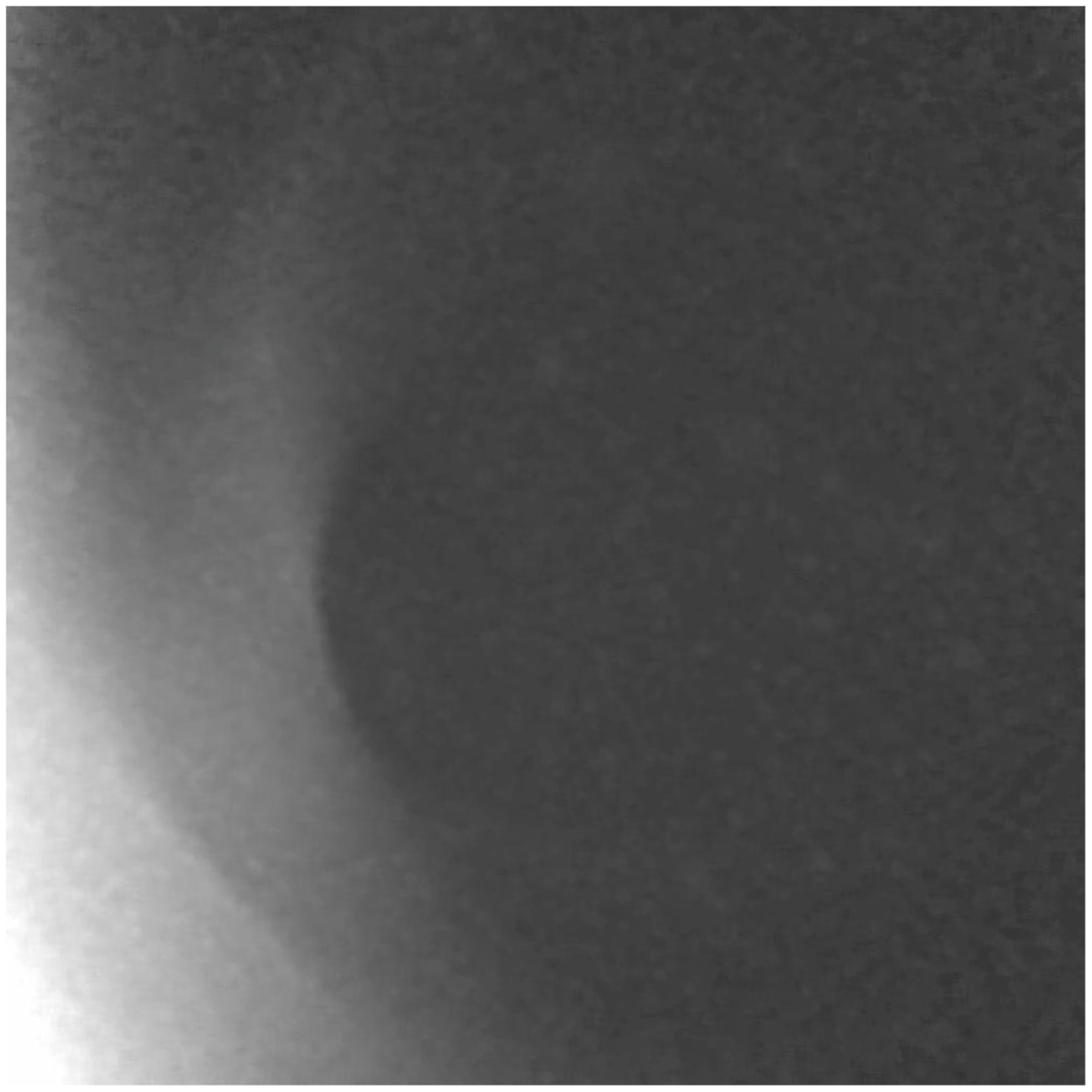} 
\end{center}
\caption{The additional stray light pattern found in the data obtained on 2022-01-24 and 2022-02-14.}
\label{newstray}
\end{figure}

We found an additional stray-light pattern in the data obtained on 2022-01-24 and 2022-02-14, as shown in Figure~\ref{newstray}.
Because this stray light appears only in the wide-band image, it cannot be removed by the v-band subtraction procedure.
The source of this additional stray light is thought to be light scattered at the inner wall of the entrance hole of the ONC-T hood,
as is indicated by the shape of the stray light (circular pattern).
For this reason, we excluded data from these two days from subsequent analyses.

\subsection{Flat-Field Correction}  \label{sec:flat}
The stray-light-subtracted image $W(x,y)-V(x,y)$ clearly shows a limb-darkening pattern (Figures~\ref{stray}c and \ref{stray}d), 
which is the same pattern as in the flat-field image (Figure~\ref{stray}e) \cite{TATSUMI2019153, KAMEDA2021-2}.
This fact means that the detector is uniformly illuminated from the front of the optics, showing that the sky brightness has certainly been detected by the ONC-T. 
We corrected this limb-darkening pattern by dividing the image by the normalized flat-field image $FLAT(x,y)$, as shown in Figure~\ref{stray}f.
Since we have not created a wide-band flat-field image, we used the v-band flat-field image instead.
The wavelength dependence of the flat-field image is negligible because the detector is identical for both bands.
The image obtained after the flat-field correction $SKY(x,y)$ is the final reduced image of the sky used for scientific analysis:
\begin{equation} SKY(x,y) = \frac{W(x,y) - V(x,y)}{FLAT(x,y)}. \end{equation}

\subsection{Sensitivity Calibration Using Stars} \label{sec:cal}
Degradation of the ONC-T sensitivity was reported after the two touchdown operations on the asteroid Ryugu \cite{KOUYAMA2021114353, Yamada2023}.
Consequently, we monitored and calibrated the sensitivity of the ONC-T in our data using the field stars in our images.
For this sensitivity calibration, we used the $W(x,y)/FLAT(x,y)$ image and not the $SKY(x,y)$ image. 
This is because the flux for the bright stars used in this calibration process in case of the $SKY(x,y)$ image is unsuitable for the sensitivity calibration 
as the v-band signal of the bright stars has been subtracted in the $SKY(x,y)$ image.
Although the $W(x,y)/FLAT(x,y)$ image includes the stray light described in Section \ref{sec:stray}, it can be removed by the aperture-photometry procedure described below. 

The sensitivity-calibration procedure for the wide-band is as follows.
First, we solved the astrometry of the images from the distribution of the stars using the astrometry-calculation code Astrometry.net \cite{Lang2010}.
Next, we matched the bright stars in the image with those in the Gaia Data Release 3 (DR3) catalog \cite{GAIA2016, GAIA2022}.
This catalog is suitable for our dataset because the wavelength coverage of Gaia's G-band, 
which is an unfiltered, white-light photometric band, is similar to that of our wide-band filter on ONC-T (Figure~\ref{filter}). 
The Gaia DR3 catalog contains around $1.806 \times 10^9$ sources, with a limiting magnitude of about $G \sim 21$~mag, 
with uncertainties of $\sim$0.3~mmag for $G<13$~mag, 1~mmag at $G=17$~mag, and 6~mmag at $G=20$~mag.
We selected stars that meet the following criteria:
\begin{enumerate}[(a)]
   \item The selected stars are in regions with stray-light intensities less than 20 DN in the $V(x,y)$ image,\label{starsela}
   \item The selected stars are in the region with normalized flat-field values greater than 0.8, and \label{starselb}
   \item The selected stars have fluxes between the 6th and 9th AB magnitude in the G-band in the Gaia DR3 catalog. \label{starselc}
\end{enumerate}
Criteria (\ref{starsela}) and (\ref{starselb}) reduce the uncertainty caused by the reduction processes of stary-light subtraction and flat-field correction.
The fraction of the area satisfying both criteria (\ref{starsela}) and (\ref{starselb}) is approximately 44\% of the total detector area in the central region of the detector.
Criterion (\ref{starselc}) reduces the uncertainty in the photometry by selecting stars that have sufficient signal but are not saturated.
The ONC-T detector is known to be linear up to $\sim$3000~DN, with $<1$\% deviation \cite{TATSUMI2019153},
and the signal value of even the brightest pixel in an image of a 6th-magnitude star is approximately within this range.

\begin{figure}
\begin{center}
\includegraphics[width=7.5cm]{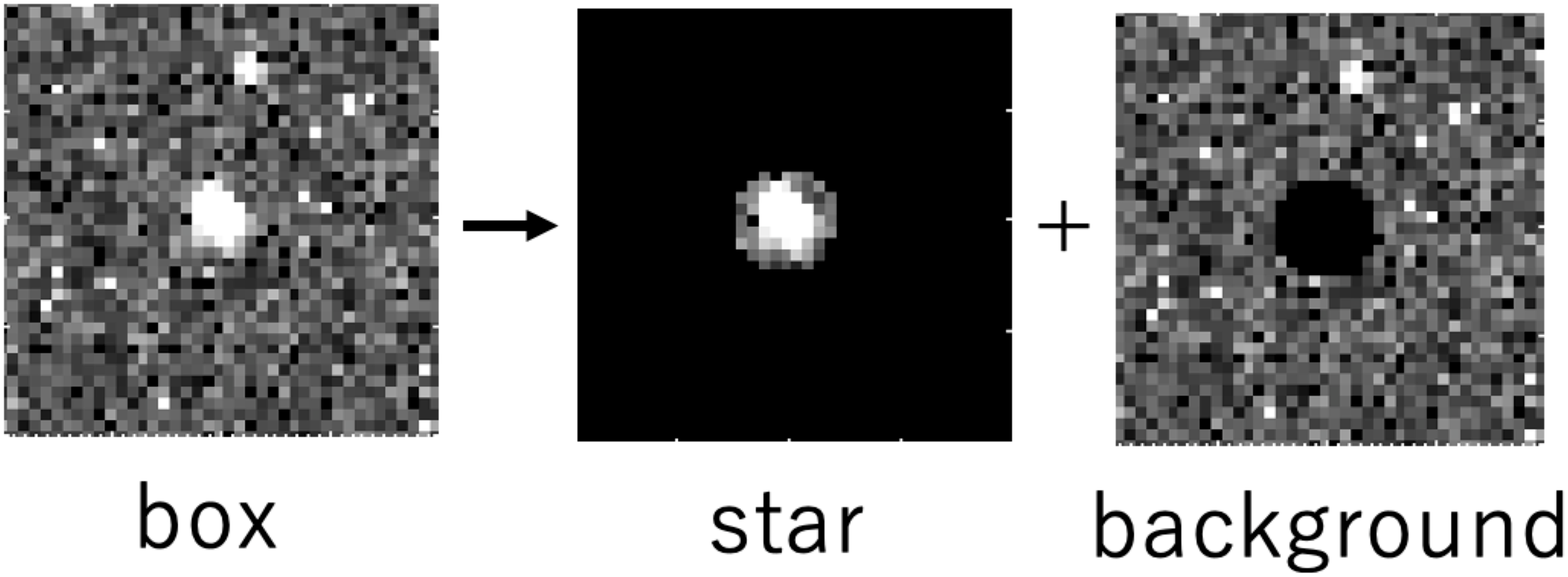} 
\end{center}
\caption{Aperture photometry of a star, as used for sensitivity calibration.  
A boxed region of $41\times41$ pixels centered on the selected star from the catalog is cut out from the $W(x,y)/FLAT(x,y)$ image (left), and it is then divided into a region containing the central star (center), which is used for the aperture photometry of the star, and the surrounding background region (right), which is used to estimate the background brightness and its noise.}
\label{aparturephoto}
\end{figure}

\begin{figure}
\begin{center}
\includegraphics[width=8cm]{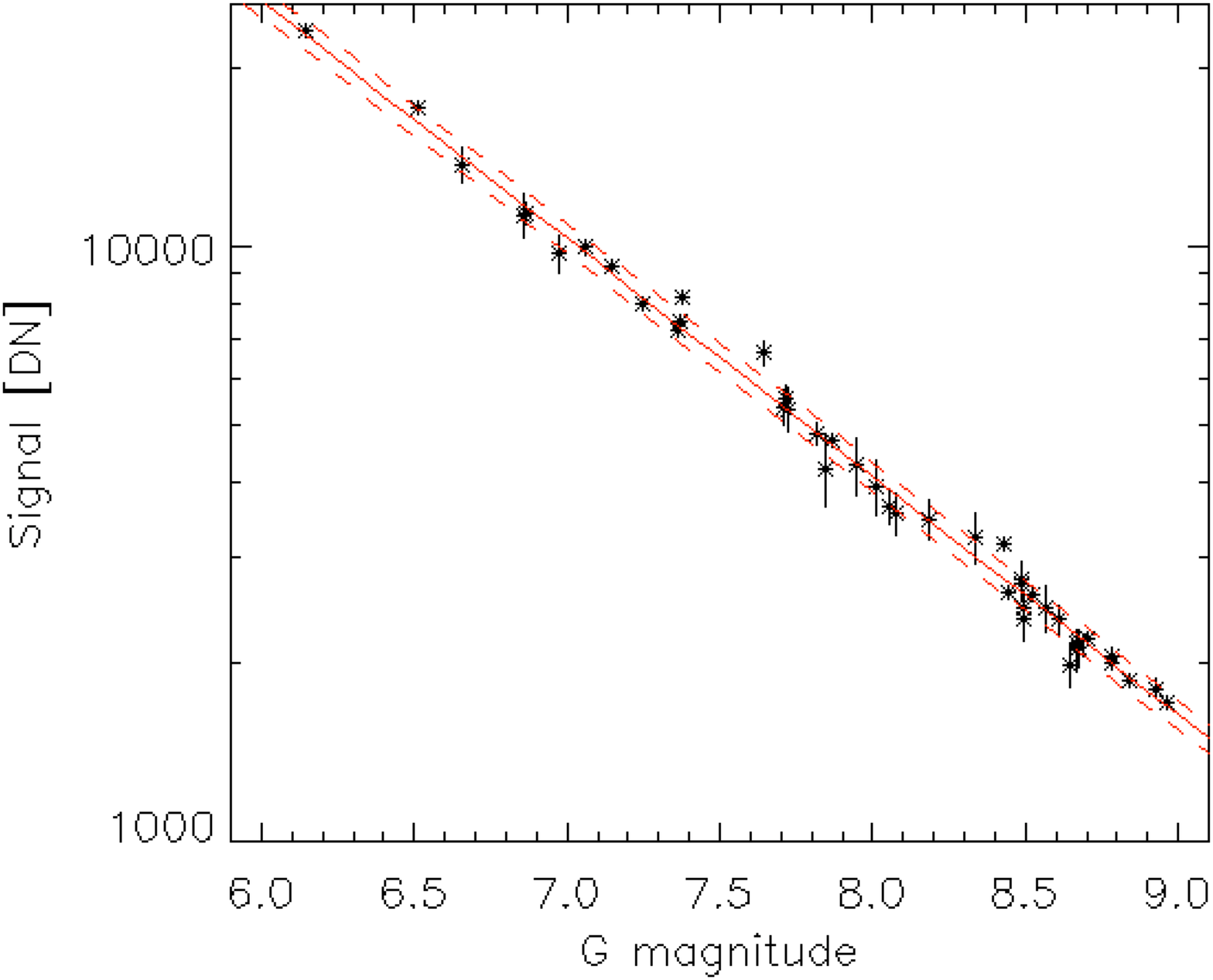} 
\end{center}
\caption{The sensitivity calibration. The linear relationship between the G-band flux of the selected stars and their detected signals in the $W(x,y)/FLAT(x,y)$ image of the 2022-12-12 wide-band data as an example of the sensitivity calibration. 
The red line shows the best fit to the data, and the dashed lines show $\pm1\sigma$ deviations.}
\label{cal1}
\end{figure}

For aperture photometry, a boxed region of $41\times41$ pixels centered on the selected star is cut out, 
and this boxed region is divided into a region centered on the star and the surrounding background region, as shown in Figure~\ref{aparturephoto}.
The radius of the circle used to cut out the star at the center is adjusted according to the G-band brightness of this star in the catalog.
Next, all pixels greater than 20~DN are masked, as they are considered to be other astronomical objects (stars and galaxies) or hot pixels caused by cosmic-ray hits. 
We applied additional masks using a $\sigma$-clipping procedure to remove the remaining bright pixels. 
Then we examined all the masked images by eye and masked any additional remaining bad pixels.
Subsequently, we calculated the background brightness and its noise by computing the average and standard deviation of the masked background region,
and we subtracted the background brightness from the star region.
We then calculated the flux from the central star by calculating the sum of the masked and background-subtracted star region and estimated its uncertainty based on the background noise.
Figure~\ref{cal1} shows the relation between the G-band fluxes of the selected stars from the catalog and their detected signal in the 2022-12-12 data.
A good linear relation between them exists in all the observed data, and we obtained the sensitivity in units of (DN/sec)/(W/m$^2$/$\mu$m/sr) by taking their ratio.

\begin{figure*}
\begin{center}
\includegraphics[width=16.5cm]{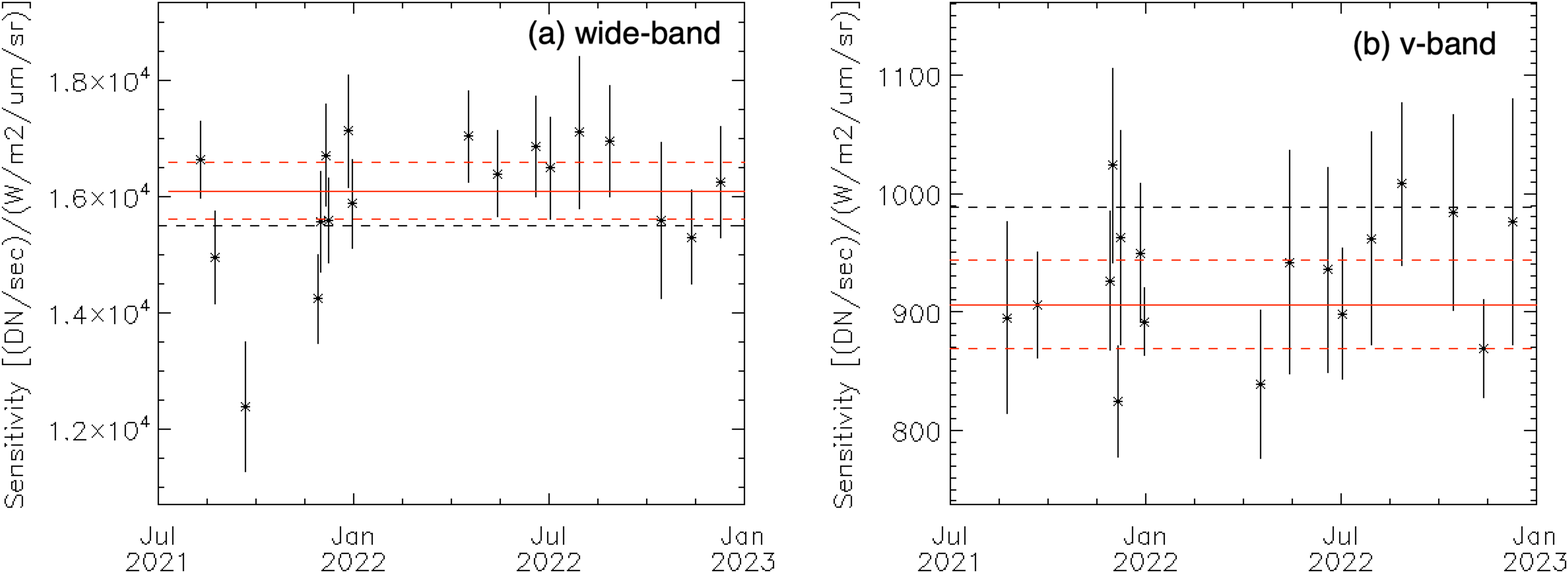} 
\end{center}
\caption{Sensitivity variation of the wide-band (left) and v-band (right) images. The red solid lines show the average of these sensitivities and the red dashed lines show their $1\sigma$ standard deviations.
The black dashed lines show the sensitivity after the return to Earth \cite{Yamada2023}.}
\label{cal2}
\end{figure*}

Figure~\ref{cal2} (left) shows the sensitivity calculated from our observed data as a function of time.
The sensitivity obtained after the return to the Earth \cite{Yamada2023} is also shown.
This figure shows that the degradation of the sensitivity has stopped and that the sensitivity has remained almost constant since the return to the Earth. 
The average and standard deviation of the sensitivity after the $\sigma$-clipping procedure is $16095 \pm 490$ (DN/sec)/(W/m$^2$/$\mu$m/sr) (red solid and dashed lines, respectively, in Figure~\ref{cal2} left), 
which we applied to all the data to obtain the sky brightness of the wide-band images. 
We treated the standard deviation of the sensitivity as a systematic uncertainty (see Section \ref{sec:heliocentric}).

We applied the same sensitivity-calibration procedure to the v-band data ($V(x,y)/FLAT(x,y)$ images) to check the consistency of the result 
because the wide-band data and v-band data share the same detector.
Gaia does not have a V-band filter, but it does have a blue band (BP) and a red band (RP),
and the V-band magnitude can be estimated from the BP and RP magnitudes \cite{Riello}.
In the selection of stars for the v-band calibration, conditions (\ref{starsela}) and (\ref{starselb}) are the same as for the wide-band calibration,
and condition (\ref{starselc}) selects stars brighter than 8th mag in the converted V-band.
Figure~\ref{cal2} (right) shows the sensitivity profile of the v-band, and we confirmed that the sensitivity remained constant in our dataset. 
The sensitivity and its 1$\sigma$ uncertainty we obtained for the v-band are $906 \pm 37$ (DN/sec)/(W/m$^2$/$\mu$m/sr).

\begin{figure*}
\begin{center}
\includegraphics[width=16.5cm]{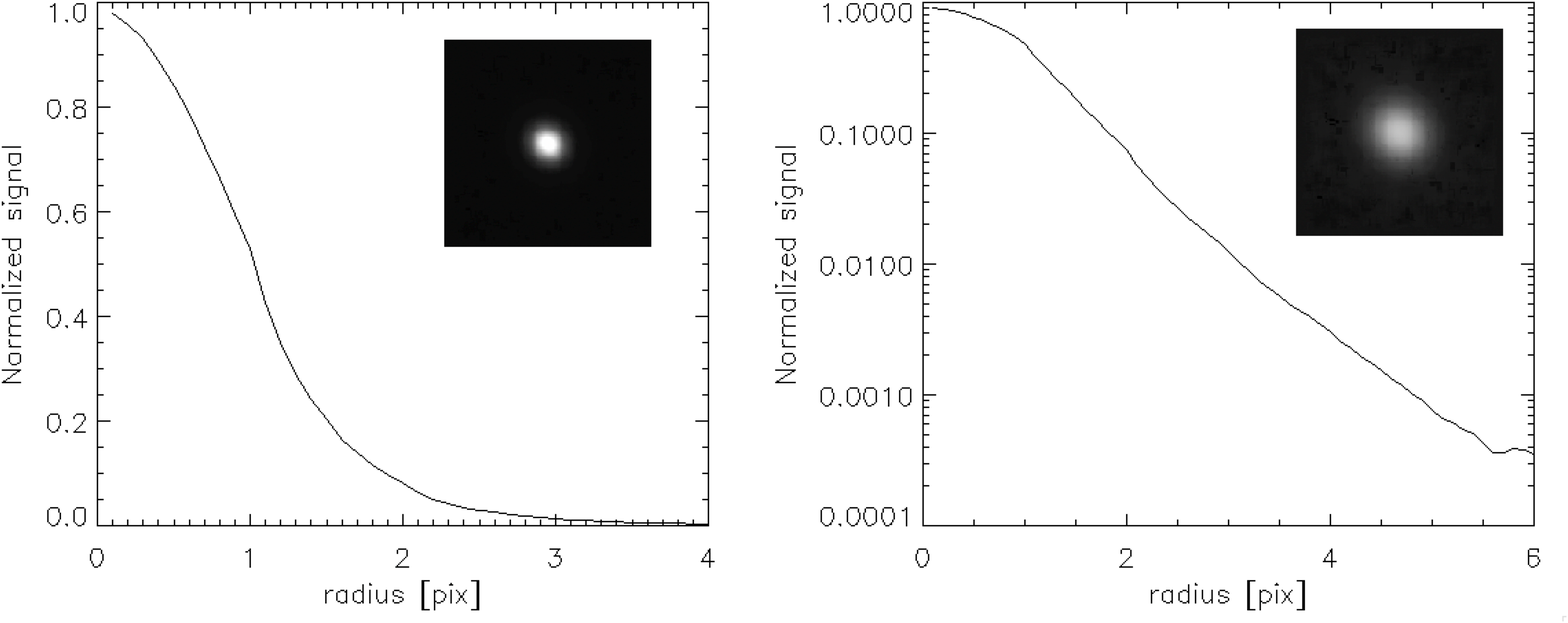} 
\end{center}
\caption{Point-Spread Function. The PSF and its radial profile with 0.1 pixel resolution are shown on a linear scale (left) and a logarithmic scale (right).}
\label{PSF}
\end{figure*}

\begin{figure*}
\begin{center}
\includegraphics[width=16.5cm]{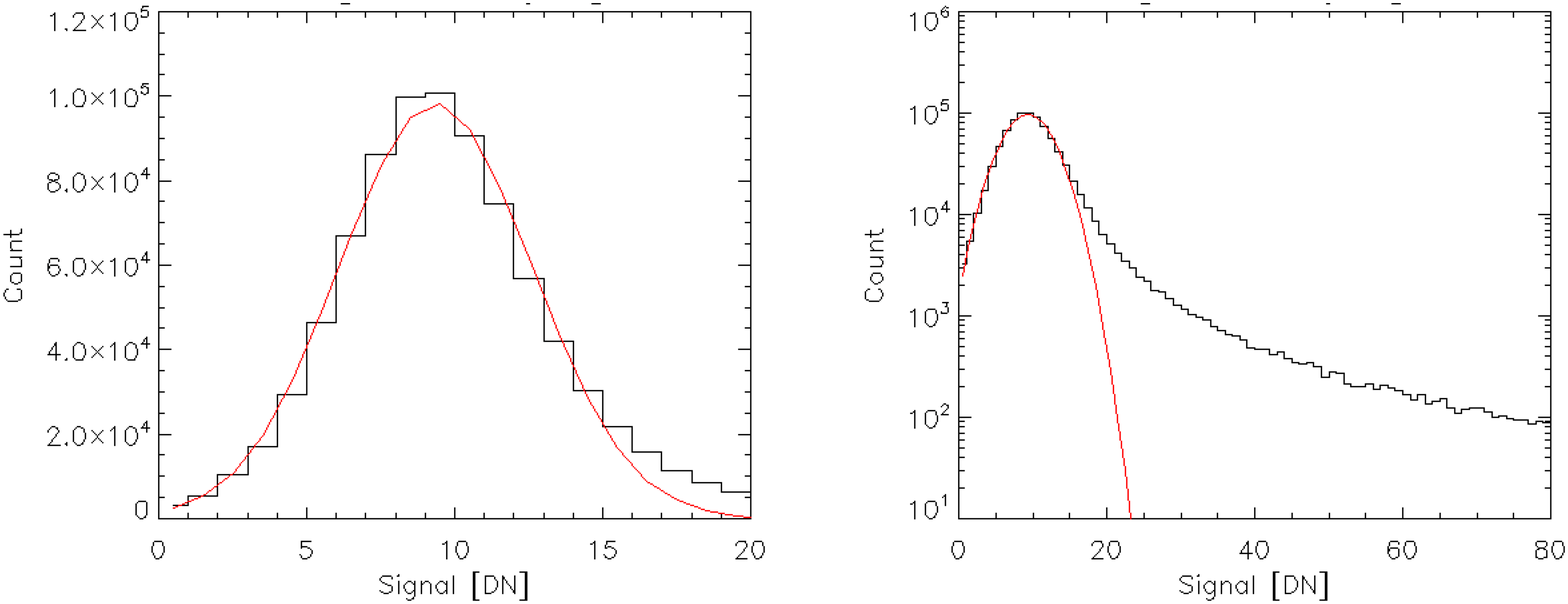} 
\end{center}
\caption{Sky brightness. A histogram of the pixel signals of the 2022-10-17 data is shown on the vertical axis with a linear scale (left) and a logarithmic scale (right). 
The red curves show the best-fit Gaussian functions}
\label{sky}
\end{figure*}

\begin{table*}
\renewcommand{\arraystretch}{1.2}
\begin{center}
\vskip 0.3cm
\caption{Obtained sky brightness and backgrounds with their uncertainties.}
\vskip -0.2cm
\begin{tabular}{lcccccc}
    \hline\noalign{\vskip3pt} 
   Date & SKY$^{(1)}$ & ISL$^{(1)}$ & DGL$^{(1)}$ & ZL$^{(1)}$ & Corr.~factor$^{(2)}$ & limmag$^{(3)}$ \\
   \hline\noalign{\vskip3pt}  
2021-08-23 &  961.1 $\pm$  23.2 $\pm$  29.3 &   28.1 $\pm$   2.2 &   34.2 $\pm$   2.6 $\pm$  27.6 &  889.3 $\pm$  16.1 $\pm$  40.3 & 1.142 & 12.79 $\pm$ 0.18 \\
2021-09-20 & 1047.8 $\pm$   6.0 $\pm$  31.9 &   34.2 $\pm$   2.7 &   33.2 $\pm$   2.5 $\pm$  25.9 &  958.4 $\pm$   8.8 $\pm$  41.1 & 1.135 & 12.80 $\pm$ 0.19 \\
2021-11-29$^{(4)}$ & 1922.1 $\pm$  12.1 $\pm$  58.6 &  299.4 $\pm$  15.1 &   46.9 $\pm$   3.3 $\pm$  30.3 & 1602.5 $\pm$  17.4 $\pm$  65.9 & 1.091 & 12.63 $\pm$ 0.16 \\
2021-11-29$^{(4)}$ & 1937.1 $\pm$  10.8 $\pm$  59.0 &  299.1 $\pm$  15.1 &   46.9 $\pm$   3.3 $\pm$  30.3 & 1619.5 $\pm$  17.3 $\pm$  66.3 & 1.091 & 12.63 $\pm$ 0.16 \\
2021-12-06$^{(4)}$ & 1864.2 $\pm$   9.8 $\pm$  56.8 &  156.1 $\pm$   8.4 &   32.5 $\pm$   2.3 $\pm$  22.1 & 1670.6 $\pm$  11.3 $\pm$  61.0 & 1.092 & 12.70 $\pm$ 0.17 \\
2021-12-06$^{(4)}$ & 1819.9 $\pm$   9.4 $\pm$  55.4 &  155.8 $\pm$   8.6 &   32.5 $\pm$   2.3 $\pm$  22.1 & 1642.8 $\pm$  11.7 $\pm$  59.7 & 1.091 & 12.70 $\pm$ 0.17 \\
2021-12-28$^{(4)}$ & 1882.5 $\pm$  10.1 $\pm$  57.4 &   45.0 $\pm$   3.1 &   24.4 $\pm$   1.8 $\pm$  18.6 & 1809.2 $\pm$   7.9 $\pm$  60.3 & 1.092 & 12.72 $\pm$ 0.17 \\
2021-12-28$^{(4)}$ & 1885.6 $\pm$  10.0 $\pm$  57.4 &   45.7 $\pm$   3.1 &   24.4 $\pm$   1.8 $\pm$  18.6 & 1770.7 $\pm$   8.2 $\pm$  60.4 & 1.092 & 12.71 $\pm$ 0.17 \\
2022-04-18 & 1599.7 $\pm$  11.0 $\pm$  48.7 &  186.3 $\pm$   8.3 &   68.7 $\pm$   5.0 $\pm$  49.1 & 1358.8 $\pm$  12.3 $\pm$  69.2 & 1.172 & 12.69 $\pm$ 0.17 \\
2022-05-16 & 1052.9 $\pm$  18.9 $\pm$  32.1 &   57.0 $\pm$   3.8 &   16.7 $\pm$   1.3 $\pm$  13.0 &  988.3 $\pm$   9.3 $\pm$  34.6 & 1.202 & 12.76 $\pm$ 0.18 \\
2022-06-20 &  857.3 $\pm$  32.7 $\pm$  26.1 &   31.8 $\pm$   2.4 &   15.1 $\pm$   1.2 $\pm$  12.2 &  806.1 $\pm$  13.9 $\pm$  28.8 & 1.198 & 12.74 $\pm$ 0.18 \\
2022-07-04 &  928.0 $\pm$   9.4 $\pm$  28.3 &   27.9 $\pm$   2.1 &   25.7 $\pm$   2.0 $\pm$  20.9 &  850.2 $\pm$   8.0 $\pm$  35.2 & 1.190 & 12.77 $\pm$ 0.18 \\
2022-08-01 &  994.2 $\pm$  15.2 $\pm$  30.3 &   31.7 $\pm$   2.4 &   17.2 $\pm$   1.3 $\pm$  13.7 &  927.5 $\pm$   8.2 $\pm$  33.2 & 1.172 & 12.79 $\pm$ 0.18 \\
2022-08-29 & 1196.4 $\pm$  12.8 $\pm$  36.4 &   38.7 $\pm$   2.7 &  132.9 $\pm$   9.9 $\pm$  99.2 & 1002.4 $\pm$  14.2 $\pm$ 105.7 & 1.147 & 12.79 $\pm$ 0.18 \\
2022-10-17 & 1984.2 $\pm$   9.4 $\pm$  60.4 &  251.7 $\pm$  12.4 &   45.2 $\pm$   3.2 $\pm$  29.6 & 1698.3 $\pm$  14.6 $\pm$  67.3 & 1.102 & 12.62 $\pm$ 0.16 \\
2022-11-14 & 1854.4 $\pm$   9.3 $\pm$  56.5 &   44.5 $\pm$   2.9 &   20.0 $\pm$   1.5 $\pm$  15.2 & 1792.9 $\pm$   7.6 $\pm$  58.5 & 1.114 & 12.69 $\pm$ 0.17 \\
2022-12-12 & 1997.9 $\pm$   7.8 $\pm$  60.9 &   37.0 $\pm$   2.6 &   23.6 $\pm$   1.8 $\pm$  18.8 & 1923.5 $\pm$   6.8 $\pm$  63.7 & 1.121 & 12.70 $\pm$ 0.17 \\
  \hline\noalign{\vskip3pt} 
  \end{tabular}
   \begin{tablenotes}
\item[1] (1) Brightness in $\lambda I_\lambda$ and its statistical and systematic uncertainty in nW/m$^2$/sr.
\item[2] (2) Correction factor used to obtain the ZL brightness toward the antisolar direction in the ecliptic plane (see Section \ref{sec:correction}).
\item[3] (3) Limiting magnitude in the G-band (see Section \ref{sec:limmag}).  
\item[4] (4) Two data sets were acquired on the same day (see Section \ref{sec:data}).  
\end{tablenotes}
\label{tab:brightness}
\end{center}
\end{table*}

\subsection{Point-Spread Function} \label{sec:psf}
We obtained a template for the point-spread function (PSF) of the ONC-T wide-band image by adding the images of several bright stars.
First, as suitable images for making the PSF template, we selected 56 objects with G-band magnitudes between 6th mag and 8th mag
that have clean stellar images, with few bad pixels or without other objects around them. 
Then, we aligned these images with 0.1~pixel resolution and added them together to obtain the PSF template.
Figure~\ref{PSF} shows the resulting PSF template and its radial profile with 0.1~pixel resolution.
The full-width half maximum (FWHM) of the PSF is 2.00 pixels, which is consistent with previous measurements \cite{KOUYAMA2021114353}.

\subsection{Sky Brightness} \label{sec:sky}
We obtained the sky brightness as the signal value of dark pixels with no stars nor hot pixels by cosmic-ray hits in the $SKY(x,y)$ image, 
using the histogram method employed to determine the dark current described in Section \ref{sec:dark}. 
We created a histogram of the $SKY(x,y)$ image, including stars and hot pixels, and obtained the best-fit Gaussian curve.
Figure~\ref{sky} shows the histogram of the $SKY(x,y)$ image and its best-fit Gaussian curve for the 2022-10-17 data.
Since the $SKY(x,y)$ image is dominated by dark pixels with no stars nor hot pixels, the peak of the pixel histogram of the $SKY(x,y)$ image represents the sky brightness.
We separate the higher signal tail of bright stars and the dark sky signals in the pixel histogram by the Gaussian fitting as shown in Figure~\ref{sky},
and we treated the 1$\sigma$ error in the peak position as the statistical uncertainty in the sky brightness.
We converted the resulting sky brightness from DN units to $\lambda I_{\lambda}$ in nW/m$^2$/sr units by applying the calibration factor obtained in Section 2.7.
Using this procedure, the systematic uncertainty in the calibration factor is transferred to a systematic uncertainty in the sky brightness.
Table \ref{tab:brightness} summarizes the obtained sky brightness and its statistical and systematic uncertainties.

At this point, the detection limit has not been determined (this is obtained in the next subsection using the sky brightness and its standard deviation). 
As it is not known how faint stars should be masked based on the star catalog, 
we determined the sky brightness by creating a histogram of the entire $SKY(x,y)$ image without masking the stars. 
This method worked well because the number of pixels observing dark sky is much larger than the number observing stars and other astronomical objects. 
Note that the stars detected in the images are masked when we determine the ZL as described in Section \ref{sec:ZL}.

\subsection{Limiting Magnitude} \label{sec:limmag}
It is important to know the limiting magnitude in our observed images because we need to mask the detected stars to derive the diffuse brightness of the sky. 
As shown in Figure~\ref{sky}, the histogram of the $SKY(x,y)$ image has a side lobe caused by the detected stars in the images, which exceeds the best-fit Gaussian curve.
The width ($\sigma$) of the Gaussian corresponds to the standard deviation of the fluctuation in the sky brightness,
and we define the limiting magnitude as the brightness of stars for which three pixels in the center of the PSF (Figure~\ref{PSF}) exceed $+2\sigma$ of the sky deviation.
We set the uncertainty in determining the limiting magnitude to be 1~DN, which is equivalent to approximately 0.2~mag uncertainty in the limiting magnitude. 
The limiting magnitude of each image and its uncertainty are summarized in Table \ref{tab:brightness}. 

\begin{figure*}
\begin{center}
\includegraphics[width=16.5cm]{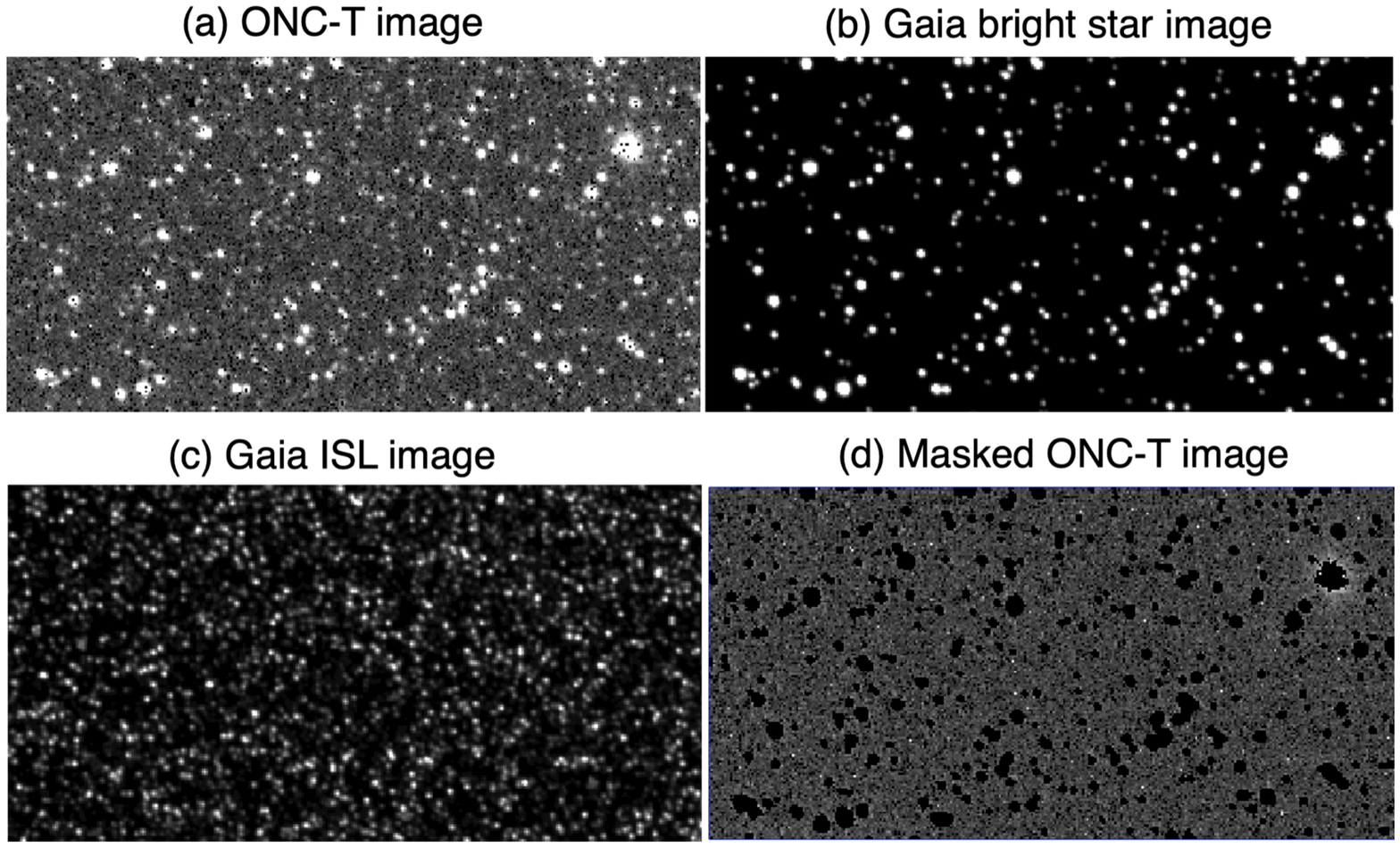} 
\end{center}
\caption{SKY(x,y) image obtained by ONC-T and ISL image. (a): A zoom-in image taken by ONC-T. 
(b): Gaia bright-star image of the same area as panel (a). We obtained this image using stars brighter than the limiting magnitude of the Gaia DR3 catalog, and we used this as a mask to conceal the stars detected in the SKY(x,y) image. 
(c): Gaia ISL image of the same area as panel (a). We constructed this image using stars fainter than the limiting magnitude of the Gaia DR3 catalog. The intensity scale of this image (c) is several times larger because the stars are too faint to be seen at the same scale. 
(d): Masked image of SKY(x,y) shown in (a) obtained using the mask image (b).}
\label{ISL}
\end{figure*}

Stars brighter than the limiting magnitude were extracted  from the Gaia DR3 catalog (see Section \ref{sec:cal}), 
convolved with the PSF (see Section \ref{sec:psf}), and distributed in the image to create a Gaia bright-star image, as shown in Figure~\ref{ISL} (b).
The distribution of stars in this Gaia bright-star image reproduces well the distribution of the stars detected in the image observed by ONC-T (Figure~\ref{ISL} (a)), 
indicating the validity of the detection limits determined by the method described above.
We used the Gaia bright-star image as a stellar mask to conceal stars when obtaining the ZL (see Section \ref{sec:ZL}).

\section{Background Subtraction}
\subsection{Integrated Starlight}\label{sec:ISL}
Stars fainter than the limiting magnitude are not detected as point sources in the observed images, 
but the sum of the light from those undetected stars, called integrated starlight (ISL), contributes to the sky brightness.
Therefore, the ISL must be estimated and subtracted from the sky brightness to obtain the ZL.

The ISL image $ISL(x,y)$ for each field is produced by stars fainter than the limiting magnitude extracted from the Gaia DR3 catalog, as shown in Figure~\ref{ISL} (c).
The average brightness of each ISL image and its $1\sigma$ statistical uncertainty are listed in Table \ref{tab:brightness}.
Since the ISL at optical wavelengths saturates when the contributions from stars down to 20th mag are added \cite{Leinert1998}, 
the depth of the Gaia DR3 catalog is sufficient.
The photometric uncertainty in the catalog is negligible compared with the uncertainty of the limiting magnitude of ONC-T.

\subsection{Diffuse Galactic Light}\label{sec:DGL}
Diffuse Galactic light (DGL) consists of starlight scattered by interstellar dust in our Galaxy \cite{Elvey1937},
and it also must be subtracted from the sky brightness to obtain the ZL.
A method commonly used to estimate the DGL is to use its correlation with the thermal emission from interstellar dust in the far infrared. 
The intensity map at $\lambda =$100~$\mu$m, which is a reprocessed composite of the COBE and IRAS maps (SFD map \cite{Schlegel1998}),
is commonly used as a template for the interstellar dust distribution. Thus,
\begin{equation} \lambda I_{DGL}  = \nu\beta_{\lambda} \cdot  d(Glat) \cdot I_{SFD}, \label{eq:SFD}\end{equation}
where $\lambda I_{DGL}$ is the DGL brightness in nW/m$^2$/sr, $I_{SFD}$ is the far-infrared intensity at 100~$\mu$m from the SFD map in MJy/sr,
$d(Glat)$  is a geometric function of the Galactic latitude ($Glat$), and $\nu\beta_{\lambda}$ is the DGL correlation factor in (nW/m$^2$/sr)/(MJy/sr).
As discussed in \cite{Sano2016b}, this geometric function is given by
\begin{equation} d(Glat) = d_0 (1-1.1g\sqrt{\sin |Glat|}), \label{eq:gb}\end{equation}
where $d_0$ is a normalizing parameter, and $g$ is the asymmetry factor of the scattering phase function \cite{Jura1979}.

This DGL correlation factor at optical and near-infrared wavelengths has been derived in many previous studies \cite{Arendt1998, Witt2008, Brandt2011, Ienaka2013, Tsumura2013b, Arai2015, Kawara2017, Onishi2018, Symons2022},
but there are two different results.
Recently, the values $\nu\beta_{\lambda} = 3.54 \pm 0.91$~(nW/m$^2$/sr)/(MJy/sr), $d_0 = 1.76$, and $g = 0.61$ have been reported from results obtained by the Long-Range Reconnaissance Imager (LORRI) on New Horizons \cite{Symons2022}.
These results were obtained at 10-50~au from the Sun, where the ZL is negligible.
The bandpass of LORRI is also similar to that of the wide-band filter of ONC-T (Figure~\ref{filter}).
However, this DGL estimate is about 5-10 times smaller than many previous results. 
For example, $\nu\beta_{\lambda} = 21.0 \pm 0.9$~[(nW/m$^2$/sr)/(MJy/sr)] and $d(Glat)=1$ (the geometric function was not considered) at 0.65~$\mu$m was found
based on observations from the Hubble Space Telescope (HST) \cite{Kawara2017}. 
In the present work, we treat the DGL estimate based on the New Horizons result as a low-level DGL estimate, 
that based on the HST result as a high-level DGL estimate, and the average of these two DGL estimates as a middle-level DGL estimate. 
We treat the difference between the low-level and the high-level DGL estimates as a systematic uncertainty.

\begin{figure*}
\begin{center}
\includegraphics[width=16.5cm]{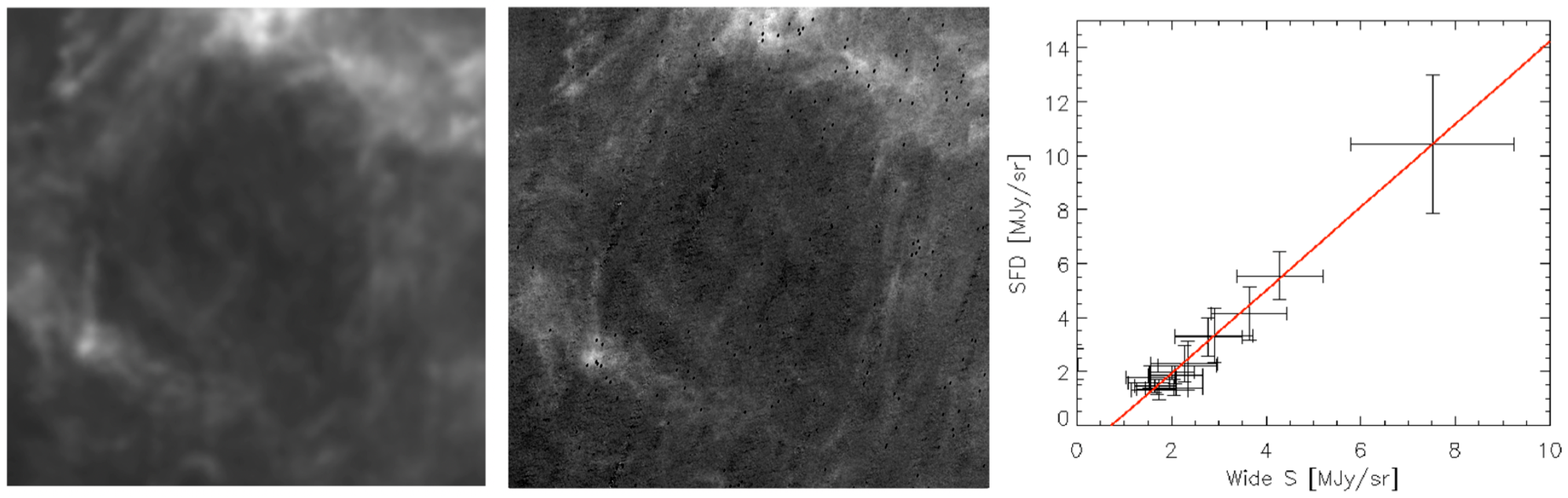} 
\end{center}
\caption{Diffuse Galactic Light. Left: DGL image for the field of 2021-08-23 based on the SFD map at $\lambda =$100~$\mu$m \cite{Schlegel1998}.
Center: DGL image for the same field based on the AKARI Wide-S map at $\lambda =$90~$\mu$m \cite{Doi2015, Takita2015} after masking the point sources.
Right: Comparison of the intensities from the SFD map and the AKARI Wide-S map of our observation fields. The red line shows the best fit to this correlation.}
\label{DGL}
\end{figure*}

The spatial resolution of the SFD map (6.1 arcmin) is insufficient relative to our data from ONC-T (22 arcsec). 
Therefore, we used a far-infrared, all-sky diffuse map based on the AKARI all-sky survey \cite{Doi2015, Takita2015} in this study.
Because these AKARI all-sky diffuse maps cover wider wavelength ranges with finer spatial resolution and better signal-to-noise ratio than the SFD map,
they can serve as a new template for the DGL estimate, replacing the SFD map.
In this study, we used the AKARI Wide-S ($\lambda = $90~$\mu$m) map, for which the spatial resolution is $\sim 1.3$ arcmin. 
One difference between these maps is that point sources have not been removed from the AKARI Wide-S map, 
whereas they have been removed from the SFD map.
Therefore, we masked all the point sources included in the AKARI Far-Infrared Bright Source Catalogue Version 2 \cite{Yamamura2018}.
Figure~\ref{DGL} compares the SFD map (left) and AKARI Wide-S map (center) in the field of 2022-08-23,
which shows that the AKARI Wide-S map has better spatial resolution than the SFD map

The publicly available AKARI Wide-S map has a ZL remainder that must be subtracted because only the smooth cloud component of the ZL has been subtracted from the raw data, 
and other ZL components, such as asteroidal dust bands, have not been subtracted \cite{Doi2015}.
The contribution from the unsubtracted ZL components is recognizable in the Wide-S map in the low-ecliptic-latitude region that we study in this work.
There is a good linear correlation between the SFD map and the AKARI Wide-S map \cite{Takita2015}, 
but the AKARI Wide-S map at low ecliptic latitudes shows deviations from the linear correlation owing to the residual ZL that remains to be subtracted.
Therefore, we used the data that other ZL components are additionally subtracted from the public AKARI Wide-S map based on a ZL asteroidal-dust-band model \cite{Ootsubo2016}.
We confirmed the good correlation between the SFD map and the additionally ZL-subtracted AKARI Wide-S map, as shown in Figure~\ref{DGL} (right).
This correlation is fitted by the equation
\begin{equation} I_{\textrm{{\small SFD}}}=  a \times I_{\textrm{{\small Wide-S}}} +c \label{eq:WideS}\end{equation}
where $I_{\textrm{{\small Wide-S}}}$ is the far-infrared intensity from the AKARI Wide-S map in MJy/sr, and $a$ and $c$ are the fitting parameters.
We obtained $a = 1.54 \pm 0.05$ and $c = -1.20 \pm 0.14$ MJy/sr in our observation fields, which are consistent with the result based on the all-sky data \cite{Doi2015, Takita2015}.
We converted the AKARI Wide-S maps in our observation fields into DGL images $DGL(x,y)$ using equations (\ref{eq:SFD}) and (\ref{eq:WideS}).

Another issue in the AKARI Wide-S map is the sky coverage. 
The AKARI all-sky map has $>99 \%$ coverage of the whole sky, but our observation fields contain regions with missing data.
We therefore used SFD data for the missing regions in the AKARI Wide-S map.
The middle-level DGL brightness based on the AKARI Wide-S map and its statistical and systematic uncertainties are listed in Table \ref{tab:brightness}.

\subsection{Extragalactic Background Light}\label{sec:EBL}
Extragalactic background light (EBL) arises from emissions integrated from the first era of star production to the present day.
Recent observations have shown that the EBL measured at optical and near-infrared wavelengths has an excess over the cumulative light from galaxies
\cite{Tsumura2013c, Matsumoto2015, Sano2015, Sano2016, Matsuura2017, Mattila2017, Zemcov2017, Lauer2022, Symons2022, Windhorst2022, Windhorst2023}, 
which means that there are unknown light sources in the universe. 
The sources for this excess are still under discussion, but some candidates that have been proposed include intra-halo light \cite{Cooray2012, Zemcov2014}, 
primordial black holes formed by the collapse of the first halos \cite{Kashlinsky2016}, the decay of hypothetical particles \cite{KOHRI2017628}, 
nearby black holes observed as faint compact objects \cite{Matsumoto2019, Matsumoto2020}, and a warm-hot intergalactic medium \cite{Zhu2022}.
We adopted $\lambda I_{EBL} = 21.98 \pm 1.83$~nW/m$^2$/sr at $\lambda$ = 0.44-0.87~$\mu$m, as observed by LORRI/New Horizons \cite{Symons2022}.
We created an EBL image $EBL(x,y)$ with all pixels having this value.

\begin{figure*}
\begin{center}
\includegraphics[width=16.5cm]{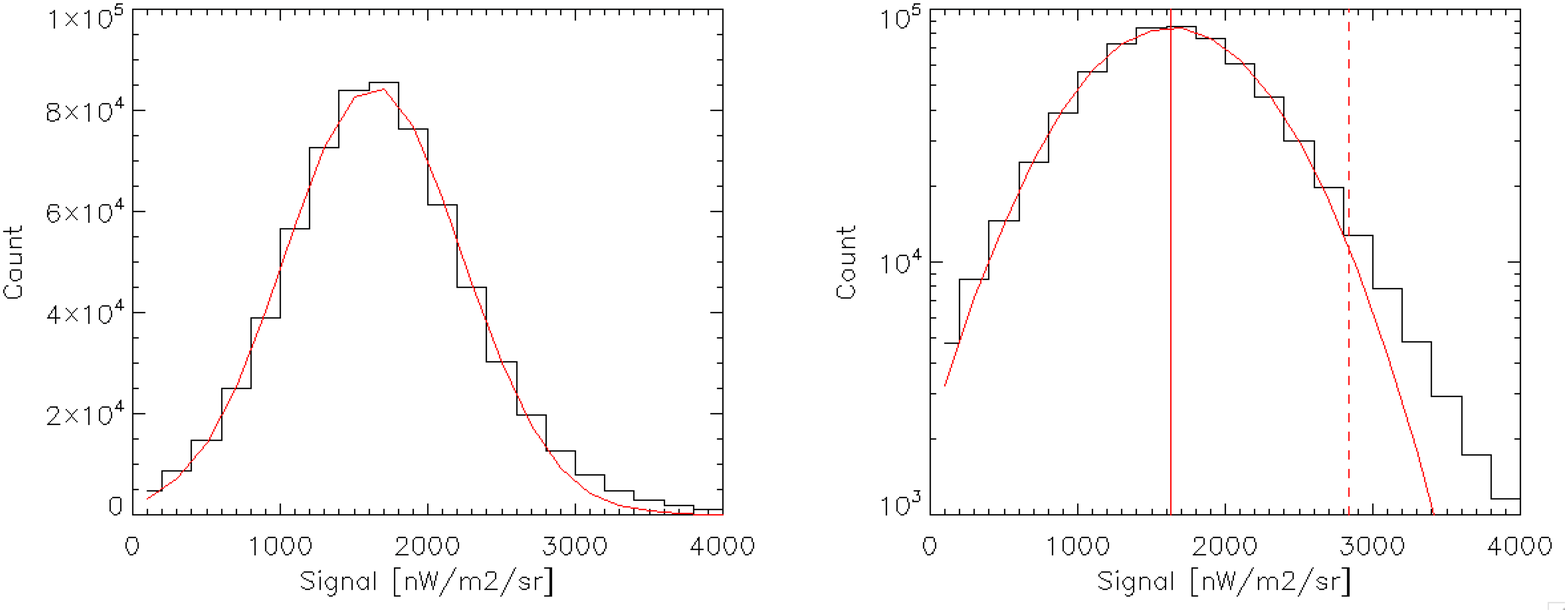} 
\end{center}
\caption{A histogram of the pixel signals from the masked ZL image of the 2022-10-17 data on a linear scale (left) and a logarithmic scale (right).
The red curves show the best-fit Gaussian functions, and the red dashed line shows the $+2\sigma$ distance from the peak position. We performed the fitting using data within this range.}
\label{ZLhist}
\end{figure*}

\subsection{Zodiacal Light} \label{sec:ZL}
The ZL is obtained by subtracting the background emissions from the observed sky brightness:
\begin{equation} ZL(x,y)  = SKY(x,y) - ISL(x,y) - DGL(x,y) - EBL(x,y). \end{equation}
We first subtracted the ISL image (Section \ref{sec:ISL}), DGL image (Section \ref{sec:DGL}), and EBL image (Section \ref{sec:EBL}) from the SKY image (Section \ref{sec:flat}) to obtain the ZL image, $ZL(x,y)$, and we masked the stars detected in the ZL image (Section \ref{sec:limmag}).
As shown in Figure~\ref{ISL} (d), this masking procedure works well for almost all stars, although the peripheries of some of the brightest stars are not masked perfectly and are smeared out. 
We then created a histogram of the area where $FLAT(x,y)$ $> 0.5$ and stray light $< 20$~DN (see Section \ref{sec:cal}) for each masked ZL image,
and we regard its peak position and its 1$\sigma$ error estimate as the brightness and statistical uncertainty of the ZL (Figure~\ref{ZLhist}).
The systematic uncertainty in the ZL comes from the systematic uncertainties in the calibration and the DGL.
Comparing the histograms of the masked ZL image (Figure~\ref{ZLhist}) with the SKY image (Figure~\ref{sky})
shows that the excess of the side lobe over the Gaussian has been reduced thanks to the stellar masking.
There is still a small excess owing to the smeared peripheries of the brightest stars and to some hot pixels caused by cosmic-ray hits, 
but this small excess has little effect on the peak position of the Gaussian 
because we performed the Gaussian fitting on data within a $+2\sigma$ range from the peak (the red dashed line in Figure~\ref{ZLhist}).
The resulting estimate of the ZL and its uncertainties are listed in Table \ref{tab:brightness} and shown in Figure~\ref{ZLbrightness} (black).

\section{Discussion}
\subsection{Absolute ZL brightness} \label{sec:ZLbrightness}
\begin{figure}
\begin{center}
\includegraphics[width=8cm]{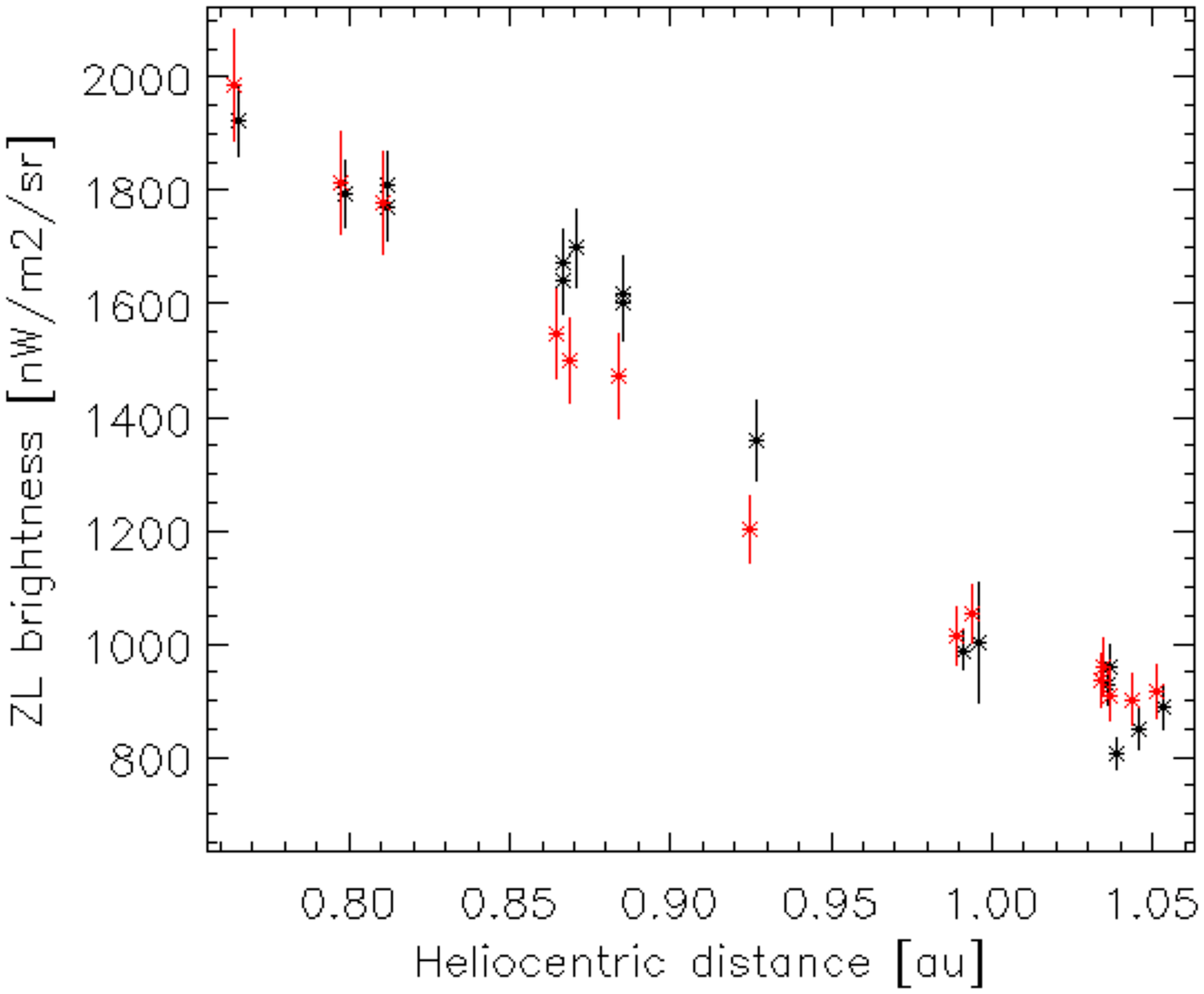} 
\end{center}
\caption{The ZL brightness of each field (black). The red points show the model brightness \cite{Kelsall1998} extrapolated to optical wavelength using the solar spectrum \cite{GUEYMARD2002443}.}
\label{ZLbrightness}
\end{figure}

We have compared the ZL brightness we observed with that predicted by the Kelsall model, 
which is based on observations of the all-sky ZL brightness obtained from COBE observations at infrared wavelengths \cite{Kelsall1998}. 
We calculated the ZL model brightness using the ZodiPy code \cite{San2022}, which implements the Kelsall model.
The shortest wavelength for which the ZL brightness can be calculated with this model is 1.25~$\mu$m, 
which is outside the range of the ONC-T wide-band data used in this study.
However, the ZL at both 1.25~$\mu$m and optical wavelengths (i.e., the ONC-T wide-band filter) results from scattered sunlight, 
and its spectral shape is the same irrespective of ecliptic latitude \cite{Tsumura2010}.
We therefore determined the ZL brightness at optical wavelengths by extrapolating from the model brightness at 1.25~$\mu$m using the solar spectrum \cite{GUEYMARD2002443}.
This extrapolation was performed using the ratio of the solar spectrum at 0.612~$\mu$m and 1.25~$\mu$m.
Since the IDP reflectance varies by about 10\% between 1.25~$\mu$m and optical wavelengths \cite{Tsumura2010, Matsuura2017, Matsumoto2018}, 
we have assumed a 10\% uncertainty in the ZL model brightness at optical wavelengths associated with this extrapolation. 
Figure~\ref{ZLbrightness} compares the observed ZL brightness with the model brightness, which shows that they are consistent with each other within the ranges of uncertainties.

\begin{figure*}
\begin{center}
\includegraphics[width=16.5cm]{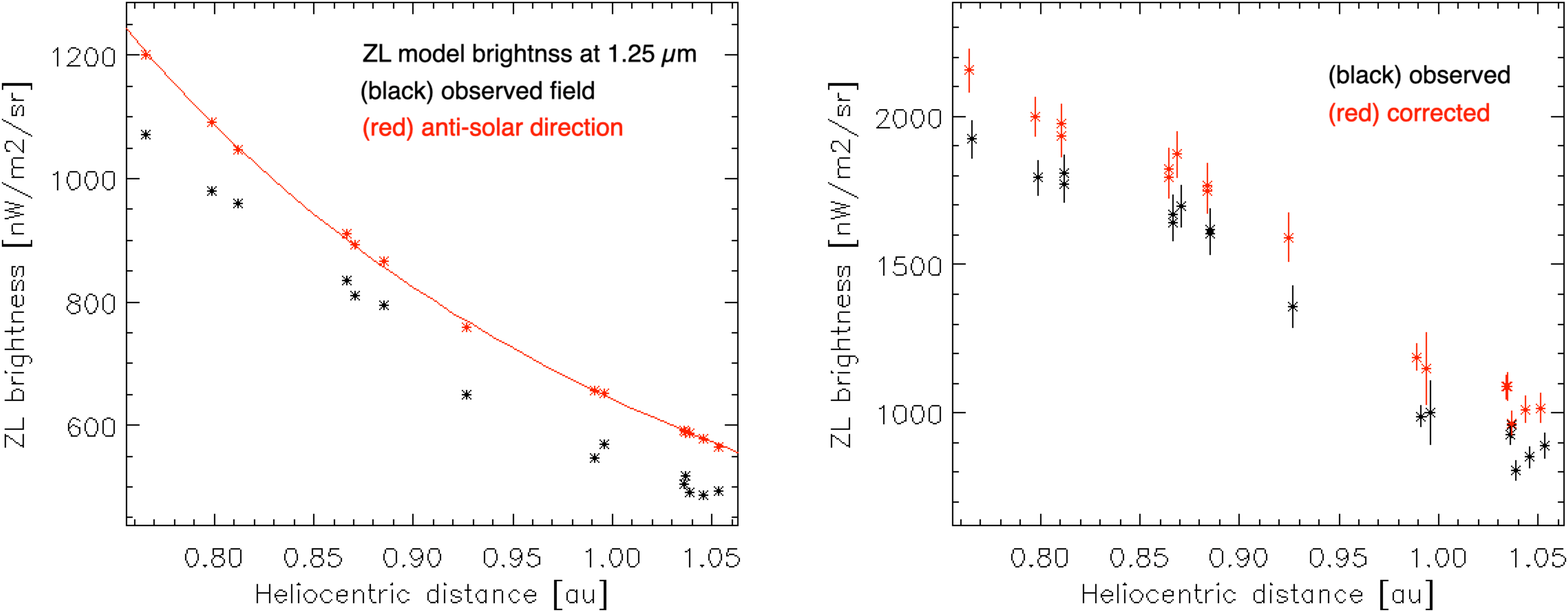} 
\end{center}
\caption{Field-variance correction.
Left: ZL model brightness of the fields observed by Hayabusa2\# (black) and toward the antisolar direction in the ecliptic plane (red) based on the Kelsall model \cite{Kelsall1998} at 1.25~$\mu$m. A radial power-law profile with $\alpha = 1.34$ is also shown. 
Right: Observed ZL brightness after subtracting the background components (black) and the corrected ZL brightness toward the antisolar direction in the ecliptic plane (red).}
\label{ZLcorrection}
\end{figure*}

Gegenschein appears in the antisolar direction, and the Kelsall model does not include it. 
However, our observed fields are shifted from the antisolar direction by $\sim$10 degrees (see Table~\ref{tab:ObsField}), and the Gegenschein is negligible there \cite{Ishiguro2013}.

There are small excesses of the observed ZL brightness over the model brightness at around 0.9~au, as shown in Figure~\ref{ZLbrightness}.
This structure may be real, but we cannot confirm this at this point due to the paucity of data points. 
Verification will require the accumulation of data from future observations.

\subsection{Field-Variance Correction} \label{sec:correction}
Since the ZL brightness measurements were obtained at different ecliptic latitudes and solar elongations,
a correction for the field variance is necessary to compare them under the same conditions in order to obtain the radial profile of the ZL.
We performed this field-variance correction based on the Kelsall model.
We calculated the seasonal average of the ZL model brightness toward the antisolar direction in the ecliptic plane at various heliocentric distances, as shown in Figure~\ref{ZLcorrection} (left, red). 
The radial power-law index of this calculated ZL model brightness is $\alpha = 1.34$ because this value is used in the Kelsall model. 
We compared this ZL brightness toward the antisolar direction with the ZL model brightness toward the fields observed by ONC-T (Figure~\ref{ZLcorrection}, left, black).
We multiplied the observed ZL brightness by the ratio of these two ZL model brightnesses at each position as correction factors in order to obtain the ZL brightness toward the antisolar direction in the ecliptic plane (Figure~\ref{ZLcorrection} right).
These correction factors are listed in Table~\ref{tab:brightness}. 

Since this correction relies on the Kelsall model with $\alpha = 1.34$, it is not self-consistent if the obtained value of $\alpha$ deviates significantly from 1.34.
As we discuss in the next subsection, however, the value of $\alpha$ we obtained from our observations is close to 1.34, so consequently this correction works.
In addition, the field variance was corrected for local brightness difference at the same elongations but different ecliptic coordinates, 
so the correction is not sensitive to $\alpha$ which represents the global distribution of IPD.

\begin{table}
\renewcommand{\arraystretch}{1.2}
\begin{center}
  \caption{Obtained radial power-law index $\alpha$ for each systematic uncertainty case.}
  \begin{tabular}{ccc}
  \hline\noalign{\vskip3pt} 
    Calibration & DGL & Value of $\alpha$ \\ 
   \hline\noalign{\vskip3pt}  
    low        & low      & $1.19 \pm 0.02$ \\
    nominal & low      & $1.22 \pm 0.02$ \\
    high       & low      & $1.21 \pm 0.02$ \\
    low        & middle & $1.29 \pm 0.02$ \\
    nominal & middle & $1.28 \pm 0.02$ \\
    high       & middle & $1.31 \pm 0.02$ \\
    low        & high     & $1.38 \pm 0.02$ \\
    nominal & high     & $1.37 \pm 0.02$ \\
    high      & high      & $1.42 \pm 0.02$ \\
  \hline\noalign{\vskip3pt} 
  \end{tabular}
  \label{tab:alpha_case}
\end{center}
\end{table}

\begin{figure*}
\begin{center}
\includegraphics[width=16.5cm]{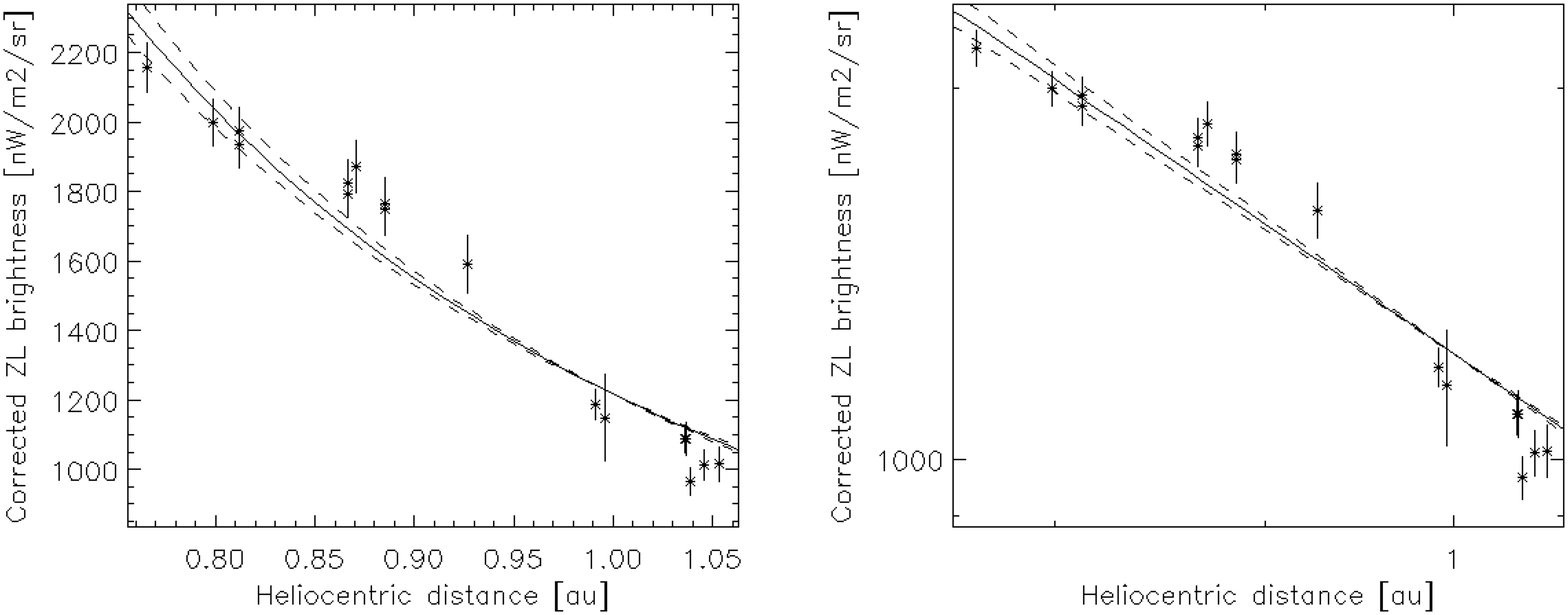} 
\end{center}
\caption{The radial profile of the ZL observed by Hayabusa2\# on 2021-2022 at 0.76-1.06~au on a linear scale (left) and a logarithmic scale (right).
The error bars include both statistical and systematic uncertainties. 
The solid curve shows the best-fit power-law function ($\alpha = 1.30$), and the dashed curves show the power-law functions for the low-DGL ($\alpha = 1.19$) and high-DGL ($\alpha = 1.42$) cases. }
\label{ZLradial}
\end{figure*}

\subsection{Dependence on Heliocentric Distance} \label{sec:heliocentric}
Based on the corrected ZL brightness, we calculated the radial power-law index $\alpha$ using the following method.
The ZL brightness has two types of uncertainties: statistical uncertainties and systematic uncertainties.
Statistical uncertainties appear randomly at each data point, while systematic uncertainties appear with a certain tendency at each data point. 
Our data contain two types of systematic uncertainties, one due to calibration (see Section \ref{sec:cal}) and the other due to DGL (see Section \ref{sec:DGL}). 
We therefore calculated the radial power-law index $\alpha$ for a total of $3 \times 3$ cases (three cases for calibration uncertainty and three cases for DGL uncertainty). 
Table \ref{tab:alpha_case} shows the values of $\alpha$ we obtained for each of these systematic-uncertainty cases.
As this table shows, the calibration uncertainty does not have a significant impact on the value of $\alpha$, since the data points only go up and down overall. 
On the other hand, the DGL uncertainty does have a significant impact on the value of $\alpha$.
From all of these values, we obtain $\alpha =  1.30 \pm 0.08$ as the final result. 
Figure~\ref{ZLradial} shows the radial profile of the ZL and the best-fit power-law function. 
Again, the excess structure at $\sim$0.9~au can be seen in Figure~\ref{ZLradial}.

\begin{table*}
\renewcommand{\arraystretch}{1.2}
\begin{center}
  \caption{Comparison of the radial power-law index $\alpha$.}
  \begin{tabular}{ccccccc}
  \hline\noalign{\vskip3pt} 
   Value of $\alpha$ & Coverage of $\alpha$ & Method & Observation wavelength & Observation site & Instrument & Reference \\ 
   \hline\noalign{\vskip3pt}  
    $1-1.5$                 & $1-3.3$ au        & ZL observation & B, R bands & Interplanetary space     & Pioneer 10/11 & \cite{Hanner1976} \\
    $1.3 \pm 0.05$     & $0.3-1$ au        & ZL observation & U, B, V bands & Interplanetary space      & Helios 1/2       & \cite{Leinert1981} \\
    $1.34 \pm 0.022$ & $>1$ au            & ZL observation & 1.25-240 $\mu$m & Geocentric orbit             & COBE             & \cite{Kelsall1998} \\
   $1.22$                   & $>1$ au            & ZL observation & 1.25-240 $\mu$m & Geocentric orbit             & COBE             & \cite{Wright1998} \\
   $1.45 \pm 0.05$    & $0.06-0.6$ au   & F-corona          &  0.5-0.9 $\mu$m & Lunar orbit                     & Clementine     & \cite{HAHN2002360} \\
   $1.59 \pm 0.02$    & $>1$ au            & ZL observation &  9 $\mu$m, 18 $\mu$m  & Geocentric orbit             & AKARI            & \cite{Kondo2016} \\
   $1.34^{(1)}$          & $1-3.3$ au         & ZL observation & B, R bands & Interplanetary space       & Pioneer 10/11 & \cite{Matsumoto2018} \\
   $1.31 - 1.35$         & $0.07-0.45$ au & F-corona          & 0.63-0.73 $\mu$m & Heliocentric orbit            & STEREO-A    & \cite{Stenborg2018} \\
   $1.31$                   & $0.1-0.4$ au     & F-corona          & 0.49-0.74 $\mu$m & Interplanetary space      & PSP                & \cite{Howard2019} \\
   $2$                        & $0.17-0.7$ au   & Dust counting   & --- & Interplanetary space      & PSP               & \cite{Szalay2020} \\
   $ 1.30 \pm 0.08$   & $0.76-1.06$ au  & ZL observation & 0.39-0.84 $\mu$m & Interplanetary space     & Hayabusa2\# & This work \\
  \hline\noalign{\vskip3pt} 
  \end{tabular}
   \begin{tablenotes}
\item[1] (1) the Kelsall model is assumed.
\end{tablenotes}
  \label{tab:alpha}
\end{center}
\end{table*}

\begin{figure}
\begin{center}
\includegraphics[width=8cm]{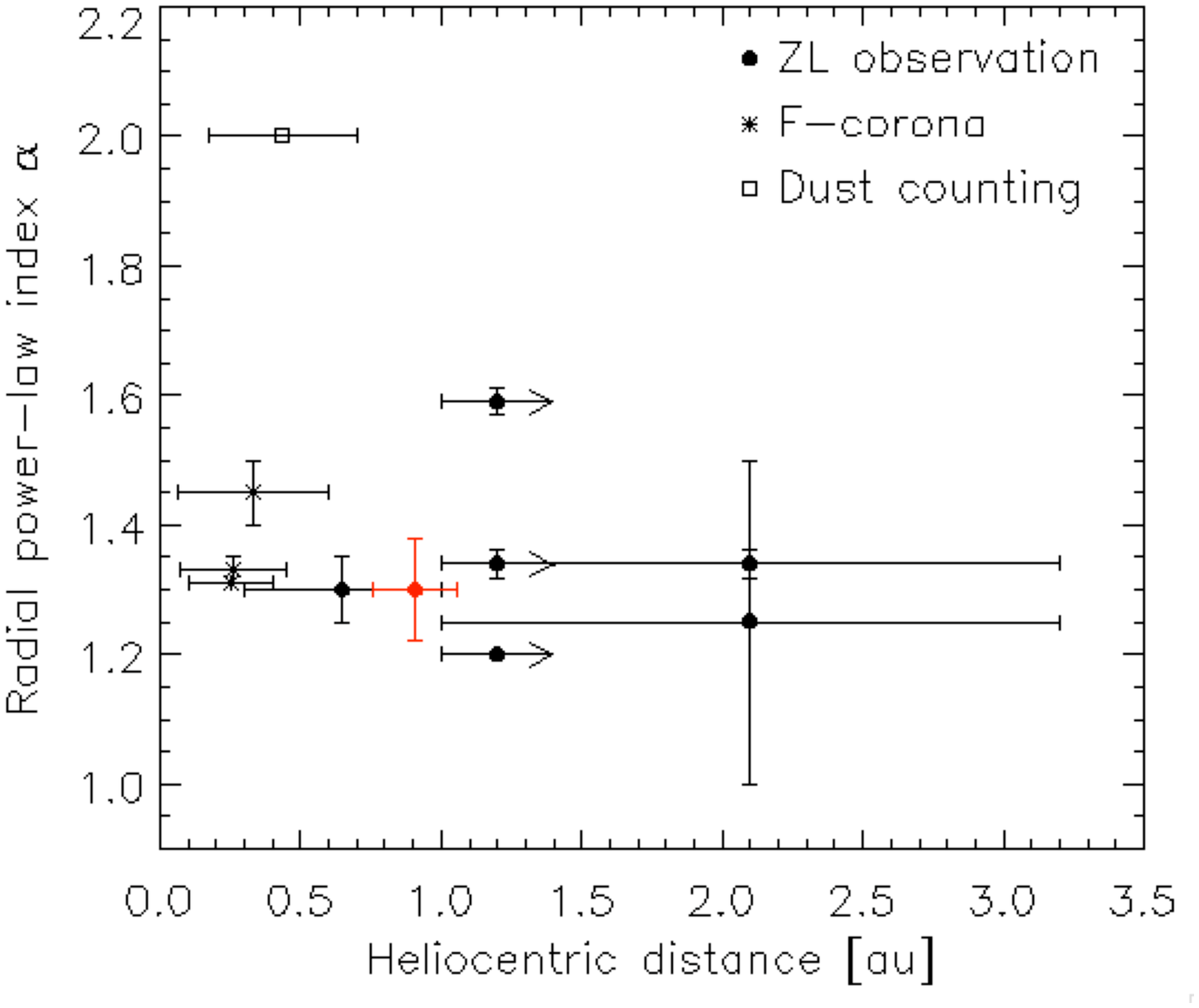} 
\end{center}
\caption{Comparison of the radial power-law index $\alpha$ as a function of heliocentric distance. The data plotted here are shown in Table \ref{tab:alpha}. The red data point shows the result from this study.}
\label{alpha}
\end{figure}

Table \ref{tab:alpha} and Figure \ref{alpha} compare the value of $\alpha$ we obtained with previous results, and they show that our result is consistent with them.
Since it is difficult to determine the radial profile of the IDP density from ZL observations obtained in a geocentric orbit or from F-corona observations, 
direct observations of the radial profile of the ZL from interplanetary space outside Earth's orbit are more reliable for this purpose.
Our observations are the first successful observations of ZL from interplanetary space in the 40 years since Helios 1/2 and Pioneer 10/11.
In addition, the ZL intensity varies with both the heliocentric distance and the solar elongation, 
both of which varied in the previous observations by Helios 1/2 and Pioneer 10/11.
On the other hand, we confined our observations to the antisolar direction (solar elongation $\sim$180~deg),
so we can impose changes in ZL brightness on changes in heliocentric distance.

Values of $\alpha$ greater than 1 are obtained from all the ZL observations, 
even though $\alpha =1$ is expected if the orbital evolution of the IDP is dominated by the PR effect \cite{BURNS19791}.
This difference is caused by dust production due to the collision of dust particles \cite{Leinert1983, GRUN1985244}, 
dust supplied by comets around 1~au \cite{Ishimoto2000}, the finiteness of the dust cloud \cite{Dijk1988}, 
or the heliocentric dependence of the local albedo of the IDP \cite{Giese1986, Levasseur1991}.
In fact, a heliocentric dependence of the local albedo of the form $r^{-0.3 \pm 0.1}$ has been reported \cite{Levasseur1991}, 
which partially explains $\alpha$ being greater than 1.

The IDPs falling into the Sun due to the PR effect decrease in size owing to evaporation, and such small IDPs are blown away by radiation pressure as $\beta$-meteoroids.
The radial profile of $\beta$-meteoroids is expected to follow a power law with $\alpha =2$,
and recent in-situ direct counting of the flux of IDPs experienced by the PSP does show $\alpha = 2$ in the range $0.17-0.7$~au \cite{Szalay2020}.
On the other hand, the reddening \cite{Tsumura2010} and polarization \cite{Takimoto2022, Takimoto2023} of the ZL spectrum indicate that the majority of IDPs seen as ZL are large ($>1$~$\mu$m);
smaller IDPs do not contribute much to the ZL even though they do exist \cite{KRUGER2014657}.
While the value $\alpha = 2$ obtained by the IDP impact-counting method is sensitive to small IDPs, 
the values of $\alpha$ between 1 and 2 obtained from ZL observations are sensitive to larger IDPs. 
The small dust particles become hotter than the larger ones \cite{Ishiguro2010},
and a hot component in the thermal emission from IDPs has been found using mid-infrared spectroscopy at $\lambda =$ 3-6~$\mu$m \cite{Ootsubo1998, OOTSUBO20002163, Hong2009, Tsumura2013a}.
We would therefore expect to obtain the value $\alpha \sim 2$ from ZL observations at $\lambda =$ 3-6~$\mu$m carried out outside Earth's orbit because small dust particles with high temperatures are mainly observed in this wavelength range.

\subsection{Future Observations}
Since ZL observations at 0.7-1~au by Hayabusa2\# will continue until the Earth swing-by at the end of 2027, 
the accuracy of the results we have reported here will be improved through the accumulation of additional observational data. 
In particular, the excess structure at $\sim$0.9~au needs to be verified by accumulating data during this phase.
After the second Earth swing-by in 2028, the Hayabusa2 spacecraft will fly to an orbit in the 1-1.5~au range \cite{MIMASU2022557}, 
so we will be able to obtain the radial profile in the outer regions of the Solar System.

Before this orbital change of Hayabusa2\#, we will have a chance to observe the radial profile of the ZL in the 1-1.5~au range from the Martian Moons eXploration (MMX) spacecraft, 
which is a Japanese sample-return mission from the Martian satellite Phobos \cite{Kuramoto2022}.
The MMX is scheduled for launch in 2024 and arrival at Mars in 2025, and we are proposing to conduct simultaneous multi-wavelength (350-1000~nm) ZL observations during this cruising phase using the Optical RadiOmeter composed of Chromatic Imagers (OROCHI) onboard MMX \cite{Kameda2021}.

In addition, we are developing an EXo-Zodiacal Infrared Telescope (EXZIT), with the aim of installing it on a spacecraft to Jupiter or farther \cite{MATUURA2014, Sano2020}.
If this instrument can be realized, we will be able to observe the radial profile of the ZL at 1-5~au as well as the EBL without the ZL foreground above 3~au.
We are also considering adding mid-infrared capabilities to EXZIT, which would allow us to examine our prediction of the $\alpha \sim 2$ index for the ZL radial profile owing to small particles.

In-situ direct dust counting is also important for comprehending the IDP distribution in the Solar System 
because it provides independent estimates of the radial variation of the IDP density. 
More quantitative comparisons between the ZL observations and in-situ direct dust counting are envisioned for future projects,
which will provide us with the differences in the distributions according to dust size and parent bodies.

\section{Summary}
We observed the ZL brightness at optical wavelengths at 0.76-1.06~au with ONC-T on the Hayabusa2\# mission.
We detected a small excess of the observed ZL brightness over the model brightness at around 0.9~au, 
but we cannot determine whether or not this structure is real at this stage due to the paucity of data points.
The radial power-law index we obtained is $\alpha = 1.30 \pm 0.08$, and the uncertainty in this estimate is dominated by the uncertainty due to the DGL estimate.
This result is consistent with previous results based on other ZL observations.

%%%%%%%%%%%%%%%%%%%%%%%%%%%%%%%%%%%%%%%%%%%%%%
%%                                          %%
%% Backmatter begins here                   %%
%%                                          %%
%%%%%%%%%%%%%%%%%%%%%%%%%%%%%%%%%%%%%%%%%%%%%%

\begin{backmatter}

\section*{Acknowledgements}%% if any
The Hayabusa2 spacecraft was developed and built under the leadership of the Japan Aerospace Exploration Agency (JAXA), 
with contributions from the German Aerospace Center (DLR) and the Centre National d'\'{e}tudes Spatiales (CNES), 
and in collaboration with NASA, Nagoya Univ., The Univ.~of Tokyo, National Astronomical Observatory of Japan (NAOJ), Univ.~of Aizu, Kobe Univ., and other universities, institutes, and companies in Japan.
This work has made use of the data obtained with AKARI, a JAXA project with the participation of the European Space Agency (ESA). 
This work also has made use of data from the ESA mission Gaia (\url{https://www.cosmos.esa.int/gaia}), 
processed by the Gaia Data Processing and Analysis Consortium (DPAC, \url{https://www.cosmos.esa.int/web/gaia/dpac/consortium}). 
Funding for the DPAC has been provided by national institutions, in particular, the institutions participating in the Gaia Multilateral Agreement.
This research has made use of the VizieR catalog access tool, CDS, Strasbourg, France (DOI : 10.26093/cds/vizier). 
The original description of the VizieR service was published in \cite{Ochsenbei2000}.
The authors thank Takafumi Ootsubo (NAOJ) for providing the ZL-subtracted AKARI Wide-S data,
and Teresa Symons (University of California, Irvine) and Michael Zemcov (Rochester Institute of Technology) for discussion on DGL estimate based on New Horizons.

\section*{Funding}%% if any
K.~Tsumura was supported by JSPS KAKENHI Grant Number 20H04744 and Tokyo City University Prioritized Studies.

\section*{Abbreviations}%% if any
CNES: Centre National d'\'{e}tudes Spatiales;
COBE: Cosmic Background Explorer;
DARTS: Data Archives and Transmission System;
DGL: Diffuse Galactic light;
DN: digital number;
DPAC: Data Processing and Analysis Consortium;
DR: Data release;
EBL: Extragalactic background Light;
ESA: European Space Agency;
EXZIT: Exo-Zodiacal Infrared Telescope;
FWHM: Full-width half maximum;
HST: Hubble Space Telescope;
IDP: Interplanetary dust particles;
ISL: Integrated star light;
JAXA: Japan Aerospace Exploration Agency;
LORRI: Long-Range Reconnaissance Imager;
MMX: Martian Moons eXploration;
NAOJ: National Astronomical Observatory of Japan;
ONC: Optical Navigation Camera;
OROCHI: Optical RadiOmeter composed of CHromatic Imagers;
PR effect: Poynting-Robertson effect;
PSF: Point spread function;
PSP: Parker Solar Probe;
STEREO: Solar TErrestrial RElations Observatory;
WISPER: Widefield Imager for Solar Probe inner telescope;
ZL: Zodiacal light.

\section*{Availability of data and materials}%% if any
The raw datasets of this study will be made available in the Hayabusa2 Science Data Archives on the Data Archives and Transmission System (DARTS) repository, 
at \url{https://www.darts.isas.jaxa.jp/planet/project/hayabusa2/}.
The analyzed datasets of this study are available from the corresponding author upon reasonable request.

\section*{Competing interests}
No competing interest is declared.

\section*{Authors' contributions}
K.Ts.~analyzed the data and wrote the manuscript. 
S.M., K.Sa., T.I., and K.Ta. contributed to the data analysis. 
M.Y., T.M., T.K., M.H., Y.Y., E.T., M.M., N.S., R.H., S.K., H.Su., K.Y., Y.C., K.O., K.Sh., H.Sa., and S.S.~contributed to ONC data acquisitions and reduction. 
M.Y.~contributed to preparing the operation sequence for the ZL observations. 
All authors discussed the results and approved the final manuscript.

%%%%%%%%%%%%%%%%%%%%%%%%%%%%%%%%%%%%%%%%%%%%%%%%%%%%%%%%%%%%%
%%                  The Bibliography                       %%
%%                                                         %%
%%  Bmc_mathpys.bst  will be used to                       %%
%%  create a .BBL file for submission.                     %%
%%  After submission of the .TEX file,                     %%
%%  you will be prompted to submit your .BBL file.         %%
%%                                                         %%
%%                                                         %%
%%  Note that the displayed Bibliography will not          %%
%%  necessarily be rendered by Latex exactly as specified  %%
%%  in the online Instructions for Authors.                %%
%%                                                         %%
%%%%%%%%%%%%%%%%%%%%%%%%%%%%%%%%%%%%%%%%%%%%%%%%%%%%%%%%%%%%%

% if your bibliography is in bibtex format, use those commands:
\bibliographystyle{vancouver} % Style BST file (bmc-mathphys, vancouver, spbasic).
\bibliography{Hayabusa2_ZL}      % Bibliography file (usually '*.bib' )
% for author-year bibliography (bmc-mathphys or spbasic)
% a) write to bib file (bmc-mathphys only)
% @settings{label, options="nameyear"}
% b) uncomment next line
\nocite{label}

% or include bibliography directly:
% \begin{thebibliography}
% \bibitem{b1}
% \end{thebibliography}

%%%%%%%%%%%%%%%%%%%%%%%%%%%%%%%%%%%
%%                               %%
%% Figures                       %%
%%                               %%
%% NB: this is for captions and  %%
%% Titles. All graphics must be  %%
%% submitted separately and NOT  %%
%% included in the Tex document  %%
%%                               %%
%%%%%%%%%%%%%%%%%%%%%%%%%%%%%%%%%%%

%%
%% Do not use \listoffigures as most will included as separate files

%%%%%%%%%%%%%%%%%%%%%%%%%%%%%%%%%%%
%%                               %%
%% Tables                        %%
%%                               %%
%%%%%%%%%%%%%%%%%%%%%%%%%%%%%%%%%%%

%% Use of \listoftables is discouraged.
%%

%%%%%%%%%%%%%%%%%%%%%%%%%%%%%%%%%%%
%%                               %%
%% Additional Files              %%
%%                               %%
%%%%%%%%%%%%%%%%%%%%%%%%%%%%%%%%%%%

\end{backmatter}
\end{document}